\documentclass[prd,preprintnumbers,
eqsecnum,floatfix,letterpaper,superscriptaddress,nofootinbib]{revtex4}
\usepackage{amsfonts,amsmath,units,wasysym,epsfig,graphicx,verbatim,color,subfigure,graphicx}
\usepackage{amsmath}
\usepackage{amssymb}
\usepackage{amsfonts}
\usepackage{bm}
\usepackage{color}
\usepackage{float}

\def\be{\begin{equation}}
\def\ee{\end{equation}}
\def\bea{\begin{eqnarray}}
\def\eea{\end{eqnarray}}

\def\e{\mathrm{e}}

\def \D{{\cal D}}
\def \Dn{{\cal D}_{\rm network}}
\def \Msun {M_\odot}
\def \N{{\cal N}}
\def \P{{\cal P}}
\def \R{{\cal R}}
\def \S{{\cal S}}
\def \T{{\cal T}}
\def \G{{\cal G}}
\def \A{{\cal A}}
\def \B{{\cal B}}
\def \Nchi {\N_{\chi^2}}
\def \H{{\cal H}}
\def \F{{\cal F}}
\def \a{\alpha}
\def \b{\beta}
\def \M{{\cal M}}
\def \fl{f_{\rm lower}}
\def \fu{f_{\rm upper}}
\def \eps{\epsilon}
\def \x{{\bf x}}
\def \y{{\bf y}}
\def \h{{\bf h}}
\def \n{{\bf n}}
\def \g{{\bf g}}
\def \s{{\bf s}}
\def \e{{\bf e}}
\def \v{{\bf v}}
\def \Dc{{\bf \Delta c}}

\def \Bom{{\bf \Omega}}
\def \fl{f_{\rm lower}}
\def \fu{f_{\rm upper}}
\def \hf {\frac{1}{2}}
\def \tn {\tilde {n}}
\def \th {\tilde {h}}
\def \no {\nonumber}
\def\lsim{\mathrel{\rlap{\lower4pt\hbox{\hskip1pt$\sim$}}
    \raise1pt\hbox{$<$}}}                
\def\gsim{\mathrel{\rlap{\lower4pt\hbox{\hskip1pt$\sim$}}
    \raise1pt\hbox{$>$}}}                

\begin{document}

\newcommand{\rhat}{\hat{r}}
\newcommand{\iotahat}{\hat{\iota}}
\newcommand{\phihat}{\hat{\phi}}

\newcommand{\IUCAA}{Inter-University Centre for Astronomy and
  Astrophysics, Post Bag 4, Ganeshkhind, Pune 411 007, India}
  
\newcommand{\IUCAAB}{Inter-University Centre for Astronomy and Astrophysics, Post Bag 4, Ganeshkhind, Pune 411 007, India}

\newcommand{\WSU}{Department of Physics \& Astronomy, Washington State University,
1245 Webster, Pullman, WA 99164-2814, U.S.A}

\newcommand{\PSU}{Institute for Gravitation and Cosmos, Physics Department, Pennsylvania State University, University Park, PA, 16802, USA}

\title{A unified approach to $\chi^2$ discriminators for searches of gravitational waves from compact binary coalescences}

\author{Sanjeev Dhurandhar}
\email{sanjeev@iucaa.in}
\affiliation{\IUCAA}

\author{Anuradha Gupta}
\email{axg645@psu.edu}
\affiliation{\PSU}
\affiliation{\IUCAA}

\author{Bhooshan Gadre}
\email{bug@iucaa.in}
\affiliation{\IUCAA}

\author{Sukanta Bose}
\email{sukanta@iucaa.in}
\affiliation{\IUCAA}
\affiliation{\WSU}

\date{\today}

\begin{abstract}

 We describe a general mathematical framework for $\chi^2$ discriminators in the context of the compact binary coalescence (CBC) search. We show that with any $\chi^2$ is associated a vector bundle over the signal manifold, that is, the manifold traced out by the signal waveforms in the function space of data segments. The $\chi^2$ is then defined as the square of the $L_2$ norm of the data vector projected onto a finite dimensional subspace (the fibre) of the Hilbert space of data trains and orthogonal to the signal waveform. Any such fibre leads to a $\chi^2$ discriminator and the full vector bundle comprising the subspaces and the base manifold constitute the $\chi^2$ discriminator. We show that the $\chi^2$ discriminators used so far in the CBC searches correspond to different fibre structures constituting different vector bundles on the same base manifold, namely, the parameter space. Several benefits accrue from this general formulation. It most importantly shows that there are a plethora of $\chi^2$s available and further gives useful insights into the vetoing procedure. It indicates procedures to formulate new $\chi^2$s that could be more effective in discriminating against commonly occurring glitches in the data. It also shows that no $\chi^2$ with a reasonable number of degrees of freedom is foolproof. It could also shed light on understanding why the traditional $\chi^2$ works so well. We show how to construct a generic $\chi^2$ given an arbitrary set of vectors in the function space of data segments. These vectors could be chosen such that glitches have maximum projection on them. Further, for glitches that can be modeled, we are able to quantify the efficiency of a given $\chi^2$ discriminator by a probability. Secondly, we propose a family of ambiguity $\chi^2$ discriminators that is an  alternative to the traditional one \cite{Allen,Allen:2005fk}. Any such ambiguity $\chi^2$ makes use of the filtered output of the template bank, thus adding negligible cost to the overall search. It is termed so because it makes significant use of the ambiguity function. We first describe the formulation with the help of Newtonian waveform, apply the ambiguity $\chi^2$ to the spinless TaylorF2 waveforms and test it on simulated data. We show that the ambiguity $\chi^2$ essentially gives a clean separation between glitches and signals. We indicate how the ambiguity $\chi^2$ can be generalized to detector networks for coherent observations. The effects of mismatch between signal and templates on a $\chi^2$ discriminator using general arguments and geometrical framework have also been investigated. 

\end{abstract}

\preprint{[LIGO-P1700206]}

\maketitle

\section{Introduction}

The detection of gravitational wave (GW) signals~\cite{Abbott:2016GW150914,Abbott:2016GW151226} intricately depends on 
comprehensively addressing the non-Gaussianity and non-stationarity of detector noise~\cite{Martynov:2016Sens} and the implementation of effective measures for discriminating noise artifacts from true signals (see, e.g., Ref.~\cite{Aasi:2014mqd}).
Non-Gaussianity and non-stationarity  
can arise from various components of the detector itself or the environment (see, e.g., Ref.~\cite{Bose:2016sqv} and the references therein). 
Matched filtering~\cite{Helstrom}, the commonly employed technique for hunting signals in noisy data, involves cross-correlating the data with a set of templates. These templates are based on gravitational waveforms that we expect the astrophysical sources of interest to emit.
But this kind of filtering, by itself, is not always sufficient to distinguish a signal from noise with high confidence. 
This is because even if a noisy feature, or ``glitch'', resembles a signal in small parts only, it can give significantly high values of the matched-filter output when it has large power.  

Addressing this problem is troublesome because there is a wide variety of noise artifacts that can produce large matched-filter output values. In spite of this difficulty, a few methods have been proposed to better discriminate signals from noise. In this work we focus on signals in ground-based detectors arising from compact binary coalescences (CBCs) involving black holes or neutron stars. These signals are transient, lasting between a fraction of a second to several minutes (see, e.g., Ref.~\cite{SD91} for their matched-filter).
The $\chi^2$ test 
described in Ref.~\cite{Allen,Allen:2005fk}, termed here as the {\em traditional} $\chi^2$ test - which we henceforth denote by $\chi^2_t$, is one such discriminator.
This test splits the broadband data into several smaller sub-bands and checks for consistency between the power expected from the signal in each sub-band with the observed power in that sub-band. This, however, is not the only type of 
discriminator that can be constructed; a few other tests have been proposed~\cite{babak2005,Hanna,HF11}, not all of which follow the $\chi^2$ distribution~\cite{Helstrom}. Moreover, while a set of $\chi^2$ tests were introduced in Ref.~\cite{HF11}, it remained to be explored whether they can be unified in a way that can facilitate construction of other $\chi^2$ tests and provide deeper insights into why and when some test is more successful than others.

In this paper, we lay down the foundations of such a formalism. We also propose a single detector and a multi-detector $\chi^2$ test and show how they and some of the tests proposed in the past fit into this unified formalism.
Let $x(t)$ describe the single-detector strain snippet providing a strong match with one of the templates of interest. When the match is above a chosen threshold value, we term the data snippet as a {\em trigger}. 
The appropriate setting for this formalism is the space of functions that describes single-detector strain of the same duration as $x(t)$, which itself belongs to the same space.
This space includes detector strains that are pure noise as well as the GW signal $h(t)$ that corresponds to the template that was triggered by $x(t)$. Detector strains which have no overlap with $h(t)$ belong to a subspace of this function space orthogonal to $h(t)$.
Ideally, noise artifacts should be detector strains which have a substantial projection on this orthogonal subspace. A $\chi^2$ discriminator is a construct that allows one to quantify how large that projection is and, thereby, helps assess how different $x(t)$ is from a GW signal.

Ingenuity and knowledge about the characteristics of glitches affecting the search are required in identifying the orthogonal subspace that has large overlaps with the glitches. On the average, this will yield large values of $\chi^2$ for the glitches, zero for the signal and an expected moderate value for Gaussian noise that is equal to the degrees of freedom of the $\chi^2$ - the dimension of the subspace. Thus, a $\chi^2$ discriminator defined on a given subspace complements another $\chi^2$ discriminator defined on a different subspace - the two together would be more effective in distinguishing, or even separating, the signals from the noise artifacts. In fact, as will be clear from the general framework for $\chi^2$ presented here, any two $\chi^2$ discriminators can be realized as a single $\chi^2$ whose orthogonal subspace is just the sum of the subspaces (in the vector-space sense) of the two individual $\chi^2$s. 
\par

This paper accomplishes two endeavors. Firstly, a general framework for $\chi^2$ discriminators is presented and shown that for each such discriminator, there is an associated vector bundle. This is in the context of a parameter space or a family of waveforms that depend on a number of parameters. It was shown in Ref.~\cite{BSD} that the family of signals has the structure of a manifold, which was called the signal manifold, with the signal parameters playing the role of coordinates. The signal manifold is in turn a subset of the function space of data trains that is in fact a Hilbert space equipped with a scalar product. A given $\chi^2$ discriminator involves a choice of a family of finite dimensional subspaces of the function space of data trains such that any such subspace is orthogonal to the signal vector belonging to  the signal manifold. Therefore, the $\chi^2$ discriminator just turns out to be the $L_2$ norm squared of the data vector projected onto this subspace. Thus, any given $\chi^2$ involves the selection of a subspaces, one at each point of the signal manifold. If the subspaces are chosen in a smooth manner, the underlying structure of the $\chi^2$ discriminator over the parameter space constitutes a smooth manifold which is a smooth vector bundle. With each $\chi^2$ discriminator there is an associated fibre structure - the base manifold of waveforms remaining the same. This is a special subcategory of vector bundles in which the base manifold remains fixed. In fibre bundle language, the $\chi^2$ discriminator can be described as a non-negative real valued function on a section of the fibre bundle - the section being defined by the vector field of projections of a given data vector onto the finite dimensional subspaces mentioned above. A framework for coherent $\chi^2$ tests has been discussed in Ref.~\cite{HF11} which has some common features with the framework presented in this paper, but here we present a more mathematically rigorous framework which has several benefits as will be discussed in the subsequent text.
\par

It is of course not necessary to choose the same dimension for the subspace at each point of the signal manifold, although it seems sensible to split the signal manifold into regions and fix the dimension of the subspace over each region, with each region having subspaces possibly of different dimensions. In this case we have a union of vector bundles instead of a single vector bundle, where the base manifold is split up into several disjoint manifolds. Since there are various ways to select such subspaces, in principle, a plethora of $\chi^2$ discriminators are possible. Thus, the important observation of this paper is that one may be able to {\it tune} the $\chi^2$ discriminator to a given family of glitches. In general, one could use different families of subspaces ($\chi^2$s) for different families of glitches.   
\par

The general framework has many benefits. It shows that there is an exceedingly rich structure yielding in principle a large number of $\chi^2$ discriminators. It indicates that how new $\chi^2$s could be formulated and be more effective in discriminating real signals with commonly occurring glitches - the probing directions for the $\chi^2$ could be taken along the glitches.
It gives valuable insights into the currently used $\chi^2$ discriminators and simplifies previous proofs. It also shows that a $\chi^2$ with reasonable degrees of freedom, say tens or few hundreds, is not foolproof. The effects of mismatch between a signal and template can be computed, in general, without reference to any specific $\chi^2$ by using general arguments.  We show how to construct a generic $\chi^2$ from an arbitrary set of vectors in the function space of data trains. These vectors could be chosen along glitches for maximum effectiveness of the $\chi^2$. Further, we are able to quantify the efficiency of a given $\chi^2$ for glitches that can be modeled as a conditional probability - we use sine-Gaussian glitches as an example.
\par

Secondly, a family of discriminators is proposed for the CBC searches. We call it the {\it ambiguity} $\chi^2$, the reason for this terminology will become clear from the text that follows. The ambiguity $\chi^2$ has very little overheads and uses essentially precomputed results, namely, the filtered output from a set of templates. In this regard it coincides very much in spirit with Hanna's construction~\cite{Hanna}. However, in many other respects differs from it. The first advantage is that it is inexpensive because the additional computations are few compared to those required for searching through the data with a bank of templates. Recently, Nitz~\cite{Nitz} has shown that the traditional $\chi^2$ can be computed in such a way that it does not add too much to the overall cost; it nonetheless involves a fair amount of computation compared to the method we propose. Our new $\chi^2$ discriminator will supplement the traditional one and, thus, help in discriminating against noise better. 
\par

The $\chi^2$ we construct is based on the following observation: If it is a signal that has triggered a given template, then the matched filter output in the vicinity of the trigger must essentially follow the ambiguity function consistent with the parameters of the trigger such as the SNR and other signal parameters. On the other hand if it is due to a noise glitch, it will not follow the ambiguity function, in general. A $\chi^2$ statistic thus can be constructed to distinguish between these two situations. The principal idea is to sample the region in the parameter space in the vicinity of the trigger template and compare the ``observed" with the ``expected" values. It is desirable to remain in the neighborhood of the trigger because the effect of the signal (more precisely the ambiguity function) would not have  died out. The $\chi^2$ discriminator's degrees of freedom to be used effectively, either the signal should have significant projection on the selected templates vectors of the $\chi^2$ or the glitch should have significant projection on the selected subspaces. Being in the neighborhood of the trigger makes sure that at least one condition is satisfied. Further, to save on computational costs, we make use of the readily available filtered output from the search pipeline which yields the observed values on a subset of the neighborhood of the trigger template. The expected values are obtained by scaling the ambiguity function by the observed SNR of the trigger template at the corresponding points of the subset. Then the differences in the expected and observed values can be used to construct a $\chi^2$ statistic. 
\par

For simplicity, we consider the spinless TaylorF2 signal waveform \cite{Buonanno} described by the two masses (or equivalently two mass parameters) and the kinematic  parameters, the time of coalescence $t_c$ and the phase at coalescence $\phi_c$. Since our method is applicable more generally (e.g., for IMRPhenomD \cite{PhnmD}), many of the formulae given here hold in general. For a demonstration of how our $\chi^2$ works, we restrict ourselves to TaylorF2 spinless waveform. Templates are constructed at $\phi_c = 0, \pi/2$ and at those discretely located mass parameters which are obtained according to a pre-decided minimal match, say $0.97$, after maximizing over $t_c$ and $\phi_c$ - this is conveniently done by defining a metric on the parameter space \cite{Owen,BSD}. Since FFT is used in the search algorithm to compute the match over a data segment to search for $t_c$, the filtered output obtained is (almost) continuous in $t_c$. Further the filtered output is obtained at $\phi_c = 0$ and $\pi/2$ and at the discretely placed templates in the mass parameters. The ambiguity $\chi^2$ makes use of tens of points in the parameter space at which the filtered output is available and which lie in the neighborhood of the trigger waveform vector. Here we present a general analysis and formulae involving both phases. For our numerical simulations and results, we consider both phases $\phi_c = 0$ and $\pi/2$. In subsection \ref{Newton} we make an exception, where for simplicity of demonstration we consider the Newtonian waveform and a single dominant phase. Without loss of generality we take the dominant phase to be $\phi_c = 0$  (if the dominant phase turns out to be $\pi/2$, the formulae can easily be adapted by interchanging the roles of $0$ and $\pi/2$ in the formulae).
\par

The paper is organized as follows: In Sec. \ref{MF} we set up the preliminaries required for our purpose and describe the matched filtering paradigm. Section \ref{FW} presents a general framework for the $\chi^2$ discriminators in terms of vector bundles. In Sec. \ref{generic} we show how a generic $\chi^2$ can be constructed from an arbitrary set of vectors by projecting out components parallel to the trigger templates and using the principal axes transformation. In Sec. \ref{ambchi2} we construct the ambiguity $\chi^2$ where we choose the test vectors from the template bank and show how the  ambiguity functions come into play. The filtered output of the search bank forms the input to the ambiguity $\chi^2$, thus adding very little to the computational cost of the search. Here we also present numerical results where the ambiguity $\chi^2$ is tested on simulated data in Sec.~\ref{ambchi2}.  We indicate how this framework can be generalized in a straight forward manner to the coherent multi-detector case~\cite{Bose:1999pj,PDB01} in Sec.~\ref{multidetector}. The effect of mismatch between the signal and the templates on the $\chi^2$ is evaluated in Sec.~\ref{msmtch}. In Section~\ref{sec:conclusions} we conclude. 
    
\section{The matched filtering paradigm}
\label{MF}

Since the matched-filter~\cite{SD91} is central to the construction of the CBC detection statistic~\cite{SD91,Allen:2005fk,Usman:2015kfa}, we briefly describe it here. Let us consider two functions, $x(t)$ and $y(t)$, defined over a data segment $[0, T]$ of duration $T$. 
When using them as vectors in the space of detector data we denote them in boldface. We define their scalar product as given below, in terms of their respective Fourier transforms $\tilde{x}(f)$ and $\tilde{y}(f)$:
\be 
 (\x ,\y) = 4 \Re \int_{\fl}^{\fu}~ df \frac{\tilde{x}^* (f) \tilde{y}(f)}{ S_{h}(f)} \,,
\label{scalar} 
\ee
where $S_h (f)$ is the one-sided power spectral-density (PSD) of the noise, which we denote by $n(t)$. The bandwidth of the detector is $[\fl, \fu]$. Often we will write $\fl = f_s$, where $f_s$ is the seismic cut-off frequency. The noise $n(t)$ is a stochastic process defined over the data segment, has an ensemble mean of zero, and is stationary in the wide sense; in the Fourier domain these properties imply that
\be
\langle \tn (f) \rangle = 0, ~~~~~ \langle \tn^* (f) \tn (f') \rangle = \hf S_h (f) \delta (f - f') \,,
\label{noise}
\ee
where angular brackets denote ensemble averages while round brackets denote scalar products. This construction makes the space of data segments a Hilbert space - a $L_2$ space with measure $d \mu \equiv df / S_h (f)$. We denote this space by $\D = L_2([0, T], \mu)$, which is a function space.  
\par

The most commonly used post-Newtonian (PN) approximant is TaylorF2, which is computed in the Fourier domain using the stationary phase approximation. We use this waveform for the signal in our discussion on the ambiguity $\chi^2$. The general form of the signal, denoted by $h$, is
\be
{\tilde h} (f) = \A ~f^{-7/6}~e^{- i \psi(f)} \,,
\ee
where the overall amplitude $\A$ depends on the binary component masses, the source distance, sky position and the relative orientation of the binary orbit to the detector. The phase $\psi (f)$ is computed to 3.5PN order explicitly~\cite{Buonanno}, and depends on 
$t_c, \phi_c$ and the mass parameters. (The sign convention for the Fourier transform in Ref.~\cite{Buonanno} is opposite to ours.) We will view these waveforms as vectors in $\D$ and denote them by the boldfaced letter $\h$.
\par

We will describe our method in terms of the Newtonian waveform, which is  simple, even if somewhat inaccurate, thereby, allowing us to draw the reader's attention more toward the intricacies of the method. However, when applying our method on simulated injections and glitches, we use PN approximants. The normalized Newtonian inspiral binary waveform in the Fourier domain is given by: 
\begin{align}
{\tilde h}(f; t_c, \tau_0, \phi_c)= \N f^{-\frac{7}{6}} e^{-i \psi_N (f; t_c, \tau_0, \phi_c)} \,,
\end{align}
where $\N$ is the normalization constant, which is determined by setting $(\h, \h) = 1$.  Thus,
\begin{align}
\N = \hf \left [\int_{\fl}^{\fu}  \frac{df}{f^{7/3}S_{h}(f)} \right ]^{- \hf} \,.
\end{align}
The phase $\psi_N (f)$ is given by:
\begin{align}
\psi_N (f; t_c, \tau_0, \phi_c)= 2 \pi f t_{c}  + \frac{6 \pi f_s \tau_0}{5} \left( \frac{f}{f_s} \right)^{-5/3} - 
\phi_c - \frac{\pi}{4}\,,
\end{align}
where $t_c$ and $\phi_c$ are the coalescence time and phase, respectively. Furthermore, we have expressed the phase in terms of a parameter more suited to this work than the chirp mass~\cite{SD91}, namely, the chirp time $\tau_0$~\cite{SD91,DS94}. The chirp times are used to construct template banks because the metric components are nearly constant in these parameters, so that the templates with a fixed minimal match cover the parameter space uniformly. Physically, $\tau_0$ is the time taken for the binary to coalesce starting from some fiducial frequency $f_{a}$. We take this fiducial frequency to be the seismic cut-off frequency $f_s = 10$ Hz for Advanced LIGO (aLIGO):
\bea
\tau_0 &=& \frac{5}{256 \pi f_s} (\pi \M f_s)^{-5/3} \, \no \\
    &\simeq& 1393 \left( \frac{f_s}{10 ~{\rm Hz}} \right)^{-8/3} \left( \frac{\M}{\Msun} \right)^{-5/3} {\rm sec} \,.
\label{chrpt}
\eea
Here $\M = \mu^{3/5} M^{2/5}$ is the chirp mass, $\mu$ and $M$ being the reduced and the total mass, respectively. $\Msun$ denotes the mass of the Sun. In terms of this normalized waveform we now define the signal and the templates. 
\par

The signal $\s$ in the data is just an amplitude $A$ multiplying the normalized waveform $\h$; thus, $\s = A \h$. The data vector, which we denote by $\x$, is then $\x = \s + \n$, when a signal is present; otherwise it is just noise, i.e., $\x = \n$, when a signal is absent. The match $c$ (correlation) is the scalar product between the data $\x$ and a (normalized) template  $\h$, that is, $c = (\x, \h)$, which is then a function of the signal parameters. In the analysis of the data for searching signals the match is maximized over signal parameters and compared with a preset threshold. In practice, for the parameters $t_c, \phi_c$, the templates need to be only defined at $\phi_c = 0$ and $\phi_c = \pi/2$, and for $t_c = 0$. This is because the search over these parameters can be done efficiently using quadratures for $\phi_c$ and the FFT algorithm for $t_c$ in a continuous fashion. The search over the mass parameters denoted by the vector parameter $\vartheta$ needs to be carried out with a finely sampled discrete bank of templates so that signals are not missed out. Thus, the search pipeline outputs the correlations $c_0$ and $c_{\pi/2}$ continuously in the parameter $t_c$~\footnote{In practice, data are sampled at a finite rate in time. Hence, the correlations are also computed at discrete values of $t_c$. However, the sampling rate is very high, and is at least twice the Nyquist frequency of the sought signal.} and at the template locations in $\vartheta$. 
\par

In the ambiguity $\chi^2$, we propose to choose our test waveforms from the template bank so that the pipeline output provides a ready-made input for computing the ambiguity $\chi^2$, which in turn means that negligible additional computational costs are involved. However, before going over to the ambiguity $\chi^2$ we will first lay down a general framework of $\chi^2$ discriminators and then show how to construct a generic $\chi^2$. The construction of the ambiguity $\chi^2$ then follows easily from the generic construction. 

\section{A general framework of the $\chi^2$ discriminators: $\chi^2$ as a vector bundle}
\label{FW}

In this section we show that there is a natural vector bundle structure that can be associated with a given 
$\chi^2$ discriminator. The $\chi^2$ then turns out to be just the square of the $L_2$ norm of the data vector projected onto the fibre of the vector bundle. 

\subsection{The mathematical structure of a $\chi^2$ discriminator}

As mentioned above, the space of data trains over a time segment $[0, T]$ is the Hilbert space $\D = L_2 ([0, T], \mu)$, where $\mu$ is a measure defined conveniently in the Fourier domain by $d \mu = df / S_h (f)$, with $S_h (f)$ being the one-sided noise PSD. The scalar product on this space is given by Eq.~(\ref{scalar}). This is essentially a $L_2$ space with the norm suitably modified. The $\chi^2$ discriminator is defined so that its value for the signal is zero and for Gaussian noise it has a $\chi^2$ distribution with a certain number of degrees of freedom. In principle, $\D$ is infinite dimensional. In practice the data are sampled at a finite but large number of points. Therefore, although $\D$ is finite dimensional, nevertheless its dimension is very large, typically, of the order of $\sim 10^6$ or more; so for all practical purposes $\D$ can be treated as infinite dimensional.
\par

Let us first pick a single waveform $\h$. Now consider the space normal to $\h$, $\Nchi (\h)$, defined by:
\be
\Nchi (\h)   = \{\x \in \D ~|~ (\x, \h) = 0 \} \,.
\ee
$\Nchi (\h)$ is infinite dimensional and isomorphic to the quotient space $\D / \{\h \}$, where $\{\h \}$ is the one dimensional subspace generated by $\h$. The $\chi^2$  test for the waveform $\h$ is defined by choosing a finite dimensional subspace of $\Nchi (\h)$, say of dimension $p$ which we denote by $\S$. This is similar to the projection operator defined in \cite{Hanna}. Then we claim that the $\chi^2$ pertaining to the waveform $\h$ is just the square of the $L_2$ norm of the data vector $\x$ projected onto $\S$. Specifically, we decompose the data vector $\x \in \D$ as,
\be
\x = \x_{\S} + \x_{\S^\perp} \,,
\ee
where $\S^{\perp}$ is the orthogonal complement of $\S$ in $\D$. $\x_{\S}$ and $\x_{\S^\perp}$ are projections of $\x$ into the subspaces $\S$ and $\S^{\perp}$, respectively. The orthogonal complement of $\S$ is defined as:
\be
\S^{\perp} = \{\x \in \D ~| ~(\x, \y) = 0, ~~~~ \forall ~~\y \in \S \} \,,
\ee
and we may write $\D$ as a direct sum of $\S$ and $\S^{\perp}$, that is, $\D = \S \oplus \S^{\perp}$.
\par

Then the statistic $\chi^2$ is,
\be
\chi^{2} (\x) = \| \x_{\S} \|^2 \,.
\ee
Thus the $\chi^2$ is defined with respect to some finite dimensional subspace $\S$ of $\D$ which is orthogonal to the waveform $\h$. Now choose any orthonormal basis in $\S$ say $\e_{\a},~~\a = 1, 2, ..., p$ so that $(\e_{\a}, \e_{\b}) = \delta_{\a \b}$, where $\delta_{\a \b}$ is the Kronecker delta. We easily verify its properties:

\begin{enumerate}

\item For a general data vector $\x \in \D$, we have:
\be
\chi^2 (\x) = \| \x_{\S} \|^2 = \sum_{\a = 1}^p |(\x, \e_{\a})|^2 \,, 
\ee   

\item Clearly, $\chi^2 (\h) = 0$, because the projection of $\h$ into the subspace $\S$ is zero or $\h_{\S} = 0$. 

\item Now let us take the noise $\n$ to be Gaussian which satisfies Eq. (\ref{noise}), i.e., $\langle \n \rangle = 0$.  Therefore, the following is valid:
\be
\chi^2 (\n) = \| \n_{\S} \|^2 = \sum_{\a = 1}^p |(\n, \e_{\a})|^2 \,.
\ee
Observe that the random variables $(\n, \e_{\a})$ are independent and Gaussian, with mean zero and variance unity because $\langle (\e_{\a}, \n) (\n, \e_{\b}) \rangle = (\e_{\a}, \e_{\b}) = \delta_{\a \b}$ (we have made use of the identity stated in Eq.~(\ref{identity})). Thus $\chi^2 (\n) $ has a $\chi^2$ distribution with $p$ degrees of freedom. 

\end{enumerate}

Therefore, the $\chi^2$ statistic satisfies the essential criteria that it is zero on the signal and is distributed  $\chi^2$ for Gaussian noise. Moreover, most importantly, the statistic is invariant under the orthogonal group in $p$ dimensions $O(p)$ acting on $\S$. That is if we choose any other orthonormal basis $\e'_{\a}$ which is related to $\e_{\a}$ by an orthogonal transformation, then the $\chi^2 (\x)$ for any vector $\x \in \D$ is invariant under this transformation. {\it This underscores the fact that it is the subspace $\S$ which is relevant, rather than any particular basis.} In order to perform the calculation, we are free to choose any orthonormal basis of $\S$. Choosing an orthonormal basis makes the statistic appear manifestly $\chi^2$ as it can be written as a sum of squares of independent Gaussian random variables with mean zero and variance unity.
\par

However, in the context of the CBC searches, we are in a more complex situation. We do not have just one waveform but a family of waveforms which depend on several parameters, such as masses, spins and other kinematical parameters. We denote these parameters by $\vartheta^a, ~~ a = 1, 2, ..., m$. As before, we may assume the waveforms to be normalized, i.e., $\| \h (\vartheta^a) \| = 1$. Then the waveforms trace out a $m$-dimensional manifold $\P$ - the signal manifold -  which is a submanifold of $\D$. Each point of $\P$ is a normalized waveform $\h (\vartheta^a)$. We now associate a $p$ dimensional subspace $\S$ orthogonal to the waveform $\h (\vartheta^a)$ at each point of $\P$ - we have a $p$ dimensional vector space ``attached" to each point of $\P$. When done in a smooth manner, this construction produces a fibre bundle with a $p$-dimensional vector space attached to each point of the $m$ dimensional manifold $\P$. The fibre bundle so obtained is a vector bundle of dimension $m + p$. We have, therefore, found a very general mathematical structure for the $\chi^2$ discriminator. Any given $\chi^2$ discriminator for a signal waveform $\h (\vartheta^a)$ is the $L_2$ norm of a given data vector $\x$ projected onto the fibre $\S$ at $\h (\vartheta^a)$. Note that here the space of data trains $\D$ has two distinct roles to play: (i) its subspaces $\S$  orthogonal to the waveforms in $\P$ form the fibres of the vector bundle, (ii) the signal manifold $\P$ is a submanifold of $\D$.

\begin{figure}[ht!]
\begin{center}
\includegraphics[width=2.5in]{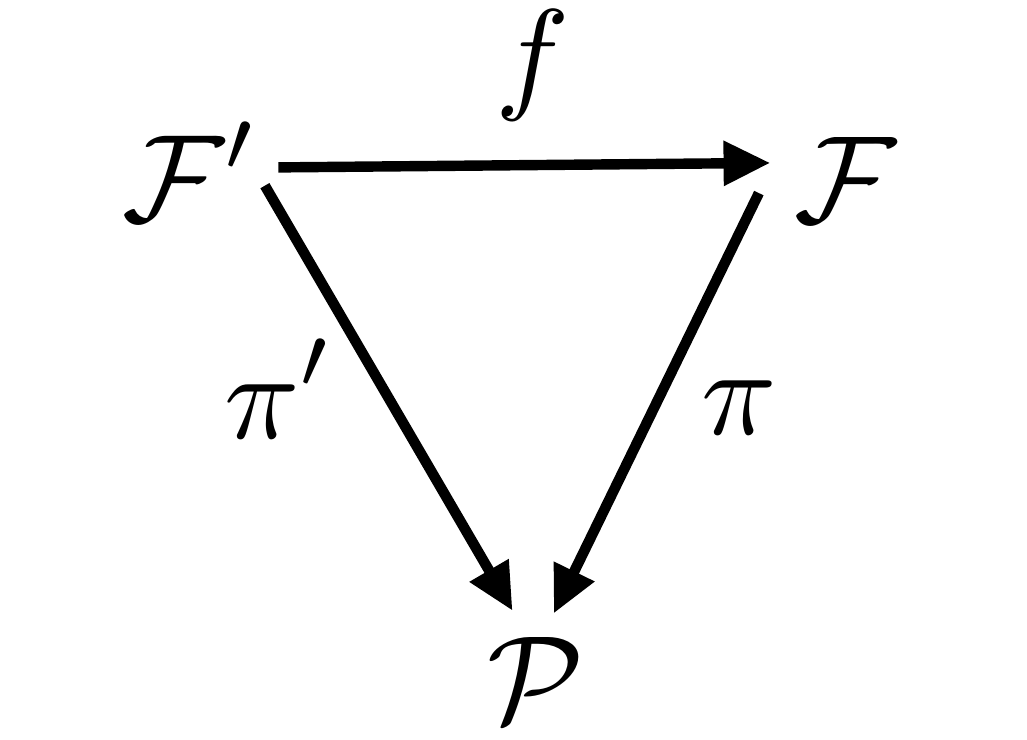}
\caption{Figure shows two vector bundles $\F$ and $\F'$ having the same base manifold $\P$ and with surjections $\pi$ and $\pi'$ respectively. $f$ is a diffeomorphism between the vector bundles such that $\pi \circ f = \pi'$. The diagram commutes.}
\label{maps} 
\end{center}
\end{figure}
\par

In the case where the signal or the base manifold has a simple topological structure, the vector bundle reduces to a product manifold - the trivial bundle. Although from the signal models considered here it may seem that the fibre bundle is always trivial, this is not the case in general. For complex signal models, for example, when the orientation of the source enters into the model in a complex manner, the base manifold could involve a spherical topological structures or when the sky location is among the search parameters, the celestial sphere could make the topological structure of the base manifold complex and a non-trivial vector bundle would  result. (For example the tangent bundle of a 2-sphere $S^2$ is non-trivial. See Ref.~\cite{Schutzbook} for an intuitive argument - one cannot comb the hair on the surface of a sphere).
\par

In effect, we have a large collection of subspaces $\S$ to choose from - in fact we have literally an infinite choice and a plethora of $\chi^2$ discriminators can be constructed. Each $\chi^2$ discriminator gives rise to a different fibre structure, the base manifold $\P$ of waveforms remaining the same. This is a special type of category of vector bundles where the base space remains fixed. We have shown this situation in Fig.~\ref{maps}, where $\pi$ and $\pi'$ are projection maps from the vector bundles $\F$ and $\F'$ respectively onto the base space $\P$. $f$ is a diffeomorphism of vector bundles and we have $\pi \circ f = \pi'$, that is, the diagram commutes. In fact $f$ induces a linear map between $\S'$ and $\S$: given a data vector $\x \in \D$ the linear map maps $\x_{\S'}$ to $\x_\S$ and therefore further induces a map between the two $\chi^2$s. In vector bundle language, each data vector $\x$ defines a section $\P_{\x}$ of the fibre bundle, namely, the vector field $\x_\S$ on $\P$ and the $\chi^2$ is a non-negative real valued function defined on $\P_{\x}$. Note however that it is not absolutely necessary to fix the dimension of $\S$ to be $p$ at each point of $\P$, but it seems convenient to do so. 

Although in principle there is enormous choice in selecting the $\chi^2$ discriminator, there are physical and practical considerations that limit the choice. Here we mention two such important considerations:

\begin{itemize}

\item The normal spaces $\S$ chosen must be such that the projection of the glitches on $\S$ is large. Specifically, if $\g$ is a glitch, then as before decomposing $\g = \g_{\S} + \g_{\S^\perp}$, we must have $\chi^2 (\g) = \chi^2 (\g_{\S}) \gg p$. This will ensure that the glitch is distinguished from the signal and Gaussian noise. This however seems to be not so simple a proposition and ingenuity may be required to select the subspaces with this property. The main problem seems to lie in modeling the glitches. On the other hand, our analysis shows that there is a lot of freedom in selecting $\S$. We expect that this facility will help find better performing discriminators.

\item It is desirable that the computational cost of evaluating the $\chi^2$ is not too large - that is the discriminator is computationally efficient. One way to achieve this, is by using precomputed scalar products such as those available from the matched filtering pipeline so that not much computational overheads are required to evaluate the $\chi^2$. The ambiguity $\chi^2$ and the bank $\chi^2$ make use of this fact.  

\end{itemize}

Finally, we point out that no $\chi^2$ statistic on a finite-dimensional manifold, even if with tens (or hundreds) of degrees of freedom, is foolproof. This is because the dimension of $\Nchi$ is so large that in a practical situation ${\rm dim} (\S^{\perp}) \gg {\rm dim} (\S)$ and there can exist glitches (vectors in $\D$) that can have very small projection in $\S$ and large projection in $\S^{\perp}$. Thus glitches can get through almost any $\chi^2$ that has a reasonable number of degrees of freedom. Generally speaking, the more the degrees of freedom a $\chi^2$ statistic possesses, that is the more directions it probes, it will tend to be more effective against glitches. This can again be seen easily from the  geometrical picture. 

\subsection{The example of the traditional $\chi^2_t$}

As a simple demonstration of the general framework described above, we show how the traditional $\chi^2_t$ fits into this framework. The main task is to describe the subspaces $\S$. We not only obtain these but also obtain an orthonormal basis for each subspace. The $\chi^2_t$, due to Allen {\it et al.}~\cite{Allen,Allen:2005fk}, is constructed in the following manner:
\par

One first partitions the detector frequency band-width into say, $p$ non-overlapping frequency sub-bands $\Delta f_1, \Delta f_2, ..., \Delta f_p$, such that the expected signal correlation in each sub-band $\Delta f_{\a}$ is the fraction $1/p$ of the full correlation $c$ of the trigger. By computing the observed signal correlation $c_{\a}$ in each sub-band and taking differences $\Delta c_{\a} = c_{\a} - c/p$ one defines the $\chi^2$ statistic as:
\be
\chi^2_t = p \sum_{\a = 1}^{p} \Delta c_{\a}^2 \,.
\ee
For a real CBC signal in Gaussian noise, this statistic is $\chi^2$ distributed with $p - 1$ degrees of freedom~\cite{Allen} when the parameters of the template perfectly match with those of the signal. For simplicity we have considered the case of known phase $\phi_c$ here. But the argument can be easily extended to the case of unknown phase. 
\par

We now exhibit how the subspace $\S$ of $p - 1$ dimensions is constructed for the traditional $\chi^2$, and how it can be imparted with an orthonormal basis in which the statistic is manifestly $\chi^2$ distributed. Let $\h$ be a normalized template waveform, that is, $\| \h \| = 1$. In the frequency domain we write it as $\th (f)$, in terms of which we can conveniently define the following $p$ waveform vectors:
\bea
\th_{\a} (f) &=& \th (f),~~~~~f \in \Delta f_{\a} \,, \no \\
          &=& 0 ~~~~~,{\rm otherwise} \,,
\eea
where $\a = 1, 2, ..., p$. These $p$ vectors are denoted by $\h_{\a}$; and they obey
$\| \h_{\a} \|^2 = 1 / p$. The waveform vectors corresponding to $\Delta c_{\a}$ can then be defined as:
\be
\Delta \h_{\a} = \h_{\a}  - \frac{\h}{p} \,.
\ee
It is easy to check that the $\Delta \h_{\a}$ are orthogonal to $\h$, that is, $(\Delta \h_{\a}, \h) = 0$. We  also observe that:
\bea
\sum_{\a = 1}^{p} \Delta \h_{\a} &=& 0 \,, \no \\
(\Delta \h_{\a}, \Delta \h_{\b}) & = & - \frac{1}{p^2}  ~~~~{\rm for}~~\a \neq \b \,.
\label{bins}
\eea
Thus the random variables $\Delta c_{\a}$ (scalar products with the data) corresponding to these waveforms are not only correlated but also algebraically dependent. But as rigorously shown in Ref.~\cite{Allen}, a $\chi^2$ statistic is obtained with $p - 1$ degrees of freedom. 
\par
Although no orthonormal basis of waveform vectors in $\S$ is explicitly given in Ref.~\cite{Allen} (there is, in fact, no need to do so), nevertheless we list one such basis here: 
\bea
\e_1 &=& \sqrt{\frac{p}{2}} (\h_1 - \h_2) \,, \no \\
\e_2 &=& \sqrt{\frac{p}{6}} (\h_1 + \h_2 - 2 \h_3) \,, \no \\
\vdots &&     \no \\
\e_{p - 1} &=& \sqrt{\frac{p}{p(p - 1)}} (\h_1 + \h_2 + ... + \h_{p - 1} - (p - 1) \h_p) \,.
\eea
The space $\S$ is spanned by the above basis vectors and is, therefore, $p - 1$ dimensional. One easily verifies that each $\e_{\a}$ is orthogonal to the waveform $\h$ and $(\e_{\a}, \e_{\b}) = \delta_{\a \b}$. As mentioned before, any orthogonal transformation of this basis will also yield another orthonormal basis. 

\subsection{Efficiency of $\chi^2$ discriminators for modeled glitches}

  In the case of glitches that can be modeled it is possible to assign an efficiency to a given $\chi^2$ discriminator. It is in fact the probability that the glitch will be removed by the $\chi^2$. To fix ideas we consider the specific case of sine-Gaussian glitches, even though our analysis can be easily generalized to other glitch models. The sine-Gaussian glitches are described in the Fourier domain as \cite{CBDL, BDGL}:
\be
g(f; A, f_0, Q) =  g_0~ \frac{A}{2 i} ~e^{- \frac{Q^2}{4 f_0^2} (f - f_0)^2} \,,
\ee
where $g_0$ is the normalization factor given by
\be
g_0^2 = \int_{\fl}^{\fu} df~  \frac{e^{- \frac{Q^2}{2 f_0^2} (f - f_0)^2}}{S_h (f)} \,,
\ee
where $A$ is the amplitude, $f_0$ the central frequency and $Q$ the quality factor. The normalization is such that 
$\|g(f; A, f_0, Q)\| = A$.
\par

Let $\G$ be the set of sine-Gaussian glitches with parameters $(A, f_0, Q)$ in some given range $\R$. Let $p(A, f_0, Q)$ be the probability density function (pdf) on $\R$ whose integral on $\R$ is unity. It is actually the relative frequency of the occurrence of such glitches in some differential volume $dA ~df_0~ dQ$ that can be estimated from the data. We can also think of $p(A, f_0, Q)$ as a prior on the set of glitches.   
\par

Let the templates be denoted by normalized waveforms $\h$, which depend on parameters, kinematic as well as dynamical. We define the set of triggers $\T$ to be:
\be
\T = \{(A, f_0, Q) \in \R~ |~~ \max (\g, \h) ~> ~\eta, ~~\g \in \G \} \,,
\ee 
where the maximum is taken over all the parameters and $\eta$ is a preset threshold.  Then $\T \subset \R$ and we define the probability,
\be
P_{\T} = \int_{\T} p (A, f_0, Q)~ dA~ df_0~ dQ < 1 \,.
\ee 
Now consider a $\chi^2$ with $p$ degrees of freedom. Then the mean of $\chi^2$ is $p$ and the standard deviation is $\sigma = \sqrt{2 p}$. We may then compute $\chi^2 (\g)$ for a trigger and if we find that this quantity is large, say, $q$ standard deviations more than the mean, i.e., $\chi^2 (\g) > p + q \sqrt{2p}$, we may decide to discard it. We may typically take $q \simeq 3$. Accordingly, we define the set:
\be
\B = \{(A, f_0, Q) \in \T ~| ~~\chi^2 (\g (A, f_0, Q) ) ~> ~p + q \sqrt{2p} \} \,,
\ee
where $\B$ is the set of triggers in $\T$ that are blocked by the $\chi^2$. The probability associated with $\B$ is
\be
P_{\B} = \int_{\B} p (A, f_0, Q)~ dA ~df_0 ~dQ \,.
\ee
Thus, the conditional probability $P_{\B} / P_{\T}$ is the probability of triggers that are blocked among all the triggers. The closer this quantity is to unity, the better the performance or the efficiency of the $\chi^2$. This would be a quantitative measure that can be used to decide on the choice of $\chi^2$ for sine-Gaussian glitches. 
\par

This analysis can be easily generalized in an analogous fashion to any other class of glitches that can be modeled. We plan to study the utility of this quantitative measure for sine-Gaussian and other modeled glitches in a future work.

\section{Constructing a generic $\chi^2$ discriminator}
\label{generic}

Using the framework described in the last section, we propose a general scheme for constructing $\chi^2$ discriminators. As we saw constructing a $\chi^2$ involves selecting subspaces $\S$ orthogonal to the signal waveforms. Further we would like these subspaces to be such that the projection of the glitches on them is as large as possible. One way to achieve this is by choosing vectors along the glitches and using them to construct a field of the subspaces $\S$ on the signal manifold. For example, we may choose the $\h_\a$ to be along sine-Gaussians~\cite{CBDL,BDGL}, or even better, choose the $\h_\a$ as a sine-Gaussian basis, if our goal is to rule out sine-Gaussians. We can do the same for other glitches if one can either model them or somehow find their directions in $\D$ and then orient the $\h_\a$ along those directions. Below we present the construct.

\subsection{Obtaining the subspaces $\S$}

Let us begin with an arbitrary set of vectors $\h_\a$, where $\a = 1, 2, ...,p$. We choose these vectors to be linearly independent (if they were linearly dependent, we can omit the dependent ones and make our set linearly independent). The $\chi^2$ statistic is constructed by taking the differences between the expected and observed correlations. The expected correlations are obtained when the data is just the signal. Consider a trigger from a template with parameters $\vartheta$ and $t_c$. We denote the trigger waveform vectors by $\h_0 (0)$ and $\h_{\pi/2} (0)$ which are time-shifted versions of the templates at $\vartheta$ by the amount $t_c$.  The corresponding correlations with the data $\x$ are denoted by $c_0 (0) = (\x, \h_0 (0))$ and $c_{\pi/2} (0) = (\x, \h_{\pi/2} (0))$. The `zero' in the trigger template is consistent with the notation that $\Delta \vartheta = \Delta t_c = 0$ for this waveform vector. The {\em observed} value of the correlation of the data $\x$ with the chosen vector 
$\h_{\a}$ is just given by:
\be
c_{\a}^{o} = (\x, \h_{\a}) \,.
\ee
We now define the expected correlations $c^{e}_{\a}$. We consider the situation of the perfect match between the signal and the trigger template. Then the signal can be written as:
\be
\s = (\s, \h_0 (0)) ~\h_0 (0) + (\s, \h_{\pi/2} (0))~ \h_{\pi/2} (0) \,,
\label{sig}
\ee
and its scalar product with the vector $\h_{\a}$ as,
\be
(\s, \h_{\a}) = (\s, \h_0 (0))~(\h_0 (0), \h_{\a}) + (\s, \h_{\pi/2} (0))~(\h_{\pi/2} (0), \h_{\a}) \,.
\label{expect}
\ee
Now the observed values of the correlations of the data with the trigger waveform vectors are:
\be
c_0 (0)  = (\x, \h_0 (0)), ~~~~ c_{\pi/2} (0)  = (\x, \h_{\pi/2} (0)) \,,
\ee
which have mean values,
\be
\langle c_0 (0) \rangle  = (\s, \h_0 (0)), ~~~~ \langle c_{\pi/2} (0) \rangle = (\s, \h_{\pi/2} (0)) \,,
\ee
because we have zero mean noise. We then replace the scalar products $(\s, \h_0 (0))$ and $(\s, \h_{\pi/2} (0))$ in Eq. (\ref{expect}) by the observed $c_0 (0) = (\x, \h_0 (0))$ and $c_{\pi/2} (0) = (\x, \h_{\pi/2} (0))$ respectively. Accordingly, we define the {\em expected} correlations as:
\be
c^{e}_{\a} = c_0 (0)~(\h_0 (0), \h_{\a}) + c_{\pi/2} (0)~(\h_{\pi/2} (0), \h_{\a})  \,,
\ee
whose mean values coincide with those given in Eq. (\ref{expect}). We then take the differences between the expected and observed correlations:
\be
\Delta c_{\a} (\x) =  c_{\a}^{o} (\x) - c^{e}_{\a} \equiv (\x, \Delta \h_{\a}) \,,
\label{delc}
\ee
where we have defined the projected vectors $\Delta \h_{\a}$ as,
\be
\Delta \h_{\a} = \h_{\a} - (\h_\a, \h_0 (0))~ \h_0 (0) - (\h_\a, \h_{\pi/2} (0))~ \h_{\pi/2} (0) \,.
\label{dhalpha}
\ee
We readily verify that $(\Delta \h_{\a}, \h_0 (0)) = (\Delta \h_{\a}, \h_{\pi/2} (0)) = 0$ for every $\a$ and therefore $\Delta \h_{\a}$ are also orthogonal to the signal $\s$. {\it Therefore $\S$ is precisely the span of $\Delta \h_{\a}$ and thus we have constructed a generic $\chi^2$ for an arbitrary collection of vectors $\h_\a$.} By taking the square of the norm of the projected vector $\x_{\S}$ we obtain the generic $\chi^2$. In the next subsection, we show how this can be done in an explicit manner by constructing an orthonormal basis for $\S$. 
\par

First we note the following properties of the linear functional $\Delta c_{\a}$: 
\be
\Delta c_{\a} (\s)  =  0, ~~~~~~~~ \langle \Delta c_{\a} (\n) \rangle = 0 \,.
\ee  
Thirdly, the $\Delta c_{\a}$ are Gaussian if the noise is Gaussian because they are linear combinations of Gaussian variables as can be seen from Eqs. (\ref{delc}) and (\ref{dhalpha}). However, we cannot right away take the sum of squares of the $\Delta c_{\a}$ to build a $\chi^2$ because (i) they are correlated and, moreover, (ii) they do not have unit variance. But these problems can be easily remedied by (i) performing a principal axes transformation and (ii) normalizing the resulting orthogonal vectors. 
To proceed with the orthogonalization, we need to compute the covariance matrix of the $\Delta c_{\a}$ which we do in the next subsection. 

\subsection{The covariance matrix, the principal axes transformation and orthonormal bases}

In order to compute the covariance matrix of $\Delta c_{\a}$ for a general $\x$, it is sufficient to consider the noise alone since $\Delta c_{\a} (\s) = 0$ and therefore $\Delta c_{\a} (\x) = \Delta c_{\a} (\n)$. Since 
$\langle \Delta c_{\a} (\n) \rangle = 0$, this computation is simple. From Eq. (\ref{delc}) we find that
\be
\Delta c_{\a} (\n) = (\n, \Delta \h_{\a}) \,.
\ee
We now make use of the following identity. For any vectors $\x, \y \in \D$ one has: 
\be
\langle (\x, \n) (\n, \y) \rangle = (\x, \y) \,.
\label{identity}
\ee
One can easily prove this identity using Eq.~(\ref{noise}). Since $\langle \Delta c_{\a} (\n) \rangle = 0$, the covariance matrix of $\Delta c_{\a}$ simplifies to:
\be
C_{\a \b} \equiv \langle \Delta c_{\a} (\n) \Delta c_{\b} (\n) \rangle \,,
\ee
One can explicitly compute this matrix using the identity given in Eq.~(\ref{identity}). We then have
\be
C_{\a \b} = (\Delta \h_{\a}, \Delta \h_{\b}) \,.
\ee
Note that, in general, the matrix $C$ is not diagonal, which implies that the $\Delta c_{\a}$ are correlated.
\par

An orthonormal set of vectors may be constructed from $\Delta \h_{\a}$. Let $O$ be the orthogonal transformation which diagonalizes the covariance matrix $C$, i.e., $O^T C O = \Lambda$, where $\Lambda$ is a diagonal matrix whose entries are the eigenvalues, $\lambda_{\a},~ \a = 1, 2, ..., p$ of $C$. 
\par
We now assert that $C$ is positive definite if $\Delta \h_{\a}$ are linearly independent (we later comment on the situation if they become linearly dependent after projecting $\h_{\a}$).  Choose any orthonormal basis $\e_{\a},~\a = 1, 2, ...,p$ of $\S$ and write $\Delta \h_{\a} = A_{\a \b} \e_{\b}$, then it is easy to see that $C_{\a \b} = A_{\a \gamma} A_{\b \gamma}$ or in matrix notation:
\be
C = AA^T  \,,
\ee
where $A$ is the matrix whose entries are $A_{\a \b}$. Then for any non-zero vector $\v \in \S$, we have $\v^T C \v = \v^T AA^T \v = (A^T \v)^T (A^T \v) > 0$. This means that $C$ is positive definite and all its eigenvalues must be positive, that is, $\lambda_{\a} > 0$ for each $\a$.
\par
Now construct,
\be
\Delta \h'_{\a} = (O^T)_{\a \b} \Delta \h_{\b} \,,
\ee 
which then satisfy,
\be
(\Delta \h'_{\a}, \Delta \h'_{\b}) = \lambda_{\a} \delta_{\a \b} \,.
\ee
The $\Delta \h'_{\a}$ are orthogonal but not orthonormal. We may then define the correlations:
\be
\Delta c_{\a}' = (\x, \Delta \h'_{\a}) \,. 
\ee
If the noise is Gaussian then $\Delta c'_{\a}$ are independent Gaussian random variables with mean zero and  variances $\lambda_{\a}$. The next step is to normalize the vectors by defining,
\be
\e_{\a} = \frac{\Delta \h'_{\a}}{\sqrt{\lambda_{\a}}} \,.
\ee
Then $\e_{\a},~ \a = 1, 2, ..., p$ form an orthonormal basis of $\S$. The $\chi^2$ then can be exhibited in terms of this basis as a sum of squares of standard normal random variables. This orthogonal transformation is not unique; any other orthogonal transformation of this basis will suffice to define the $\chi^2$ in this way. The $\chi^2$ discriminator is then defined as
\be
\chi^2 = \sum_{\a = 1}^p \frac{\Delta c_{\a}^{'2}}{\lambda_{\a}} \equiv \Delta c_{\a} [C^{-1}]^{\a \b} \Delta c_{\b} \,.
\label{final}
\ee
For Gaussian noise obeying Eq.~(\ref{noise}), the above statistic has a $\chi^2$ distribution with $p$ degrees of freedom.
\par

It may happen in certain situations that after projection the vectors $\Delta \h_\a$ may become linearly dependent or almost linearly dependent at certain points or regions of the signal manifold, or in other words the matrix $C$ becomes either singular or ill conditioned. In this case we will have some of the eigenvalues $\lambda_{\a}$ either zero or close to zero. The remedy is then to remove such eigenvalues or degrees of freedom from the $\chi^2$ and we will then nevertheless obtain a $\chi^2$ although with fewer degrees of freedom. This is acceptable if the degrees of freedom are not very much less than $p$. It may be possible to avoid such a situation by choosing the original vectors $\h_{\a}$ judiciously.   

The $\chi^2$ discriminators constructed so far in the literature, for example, the traditional $\chi^2_t$ and the ambiguity $\chi^2$ that we are going to describe in the  next  section, are special cases of this generic $\chi^2$. {\it But the main power of this procedure is that the $\h_\a$ may be chosen to be along the glitches that occur in the data}, which could yield a very powerful discriminator. This may be difficult to implement in practice because of the wide morphology of glitches, nevertheless, we may be able to do so for a particular class of glitches. This may be a direction worth following in the future. 

\section{The ambiguity $\chi^2$ discriminator}
\label{ambchi2}

 In this section we construct the ambiguity $\chi^2$ which is a special case of the generic $\chi^2$ described in the previous section. The idea is to (i) choose the vectors $\h_{\a}$ from the template bank and (ii) as far as possible from the vicinity of the trigger template. The reason for choosing the test waveform vectors from the template bank is that we can use the filtered output of the template bank as an input to a $\chi^2$ discriminator and therefore obtain a $\chi^2$ discriminator with very little additional computational cost. The construction essentially then follows the procedure we had adopted in the case of the generic $\chi^2$ in the previous section. 
\par

\subsection{The construction}

We choose several waveform vectors surrounding this trigger waveform vector. In general we will take the number of waveforms to be $p$. We choose the vectors at the points $(t_c + \Delta t_{c i}, \vartheta + \Delta \vartheta_i)$ and at any one of the phases $\phi_c = 0$ and $\phi_c = \pi/2$, where $i = 1, 2, ..., p$; for example, for $i = 1, 2, ..., p_1$, we may choose $\phi_c = 0$ and for the rest from $i = p_1 + 1, ..., p$ we may choose $\phi_c = \pi/2$. The $\Delta \vartheta_i$ are chosen such that they correspond to the templates in the bank. On the other hand, we can choose almost any value of $\Delta t_{c i}$ because the filtered output is almost continuous in the parameter $t_c$. We denote these vectors by $\h_{0 i}$ and $\h_{\pi/2 i}$ corresponding to the phases $0$ and $\pi/2$, respectively - they are time-shifted versions of the templates at $\vartheta + \Delta \vartheta_i$ by the amount $t_c + \Delta t_{c i}$. 
\par

We consider a signal whose $\vartheta$ exactly matches with the template. In general, there will be a mismatch between the signal parameters and the template parameters, but because the templates are densely spaced, the mismatch is expected to be small. We will show in Sec.~\ref{msmtch} that a small mismatch does not change the $\chi^2$ value significantly and, therefore, the test is robust against a small mismatch of parameters. This is also validated by our numerical simulations discussed in Sec.~\ref{ambmsmtch}. The waveform vectors satisfy the following orthonormality conditions:
\be
(\h_0 (0), \h_0 (0)) =  (\h_{\pi/2} (0), \h_{\pi/2} (0)) = (\h_{0 i}, \h_{0 i}) =  (\h_{\pi/2 i}, \h_{\pi/2 i} ) = 1 \,,
\ee
\be  
(\h_0 (0), \h_{\pi/2} (0)) = (\h_{0 i}, \h_{\pi/2 i}) = 0 \,.
\ee
 It is important to realize that the scalar products other than those listed above {\it need not vanish.} That is, in general, $(\h_{0 i}, \h_{0 j}) \neq 0$ or $(\h_{\pi/2 i}, \h_{\pi/2 j}) \neq 0$, etc., for any $i$ or $j$. Similarly, scalar products of these waveform vectors with the trigger data vectors need not vanish. This has the important consequence that the scalar products of the data with the waveform vectors can lead to random variables having non-vanishing covariances. This fact is taken into consideration when constructing the ambiguity $\chi^2$ statistic. Referring back to Sec.~\ref{generic}, these are precisely the vectors $\h_{\a}$ pertaining to the generic $\chi^2$, and we may therefore denote the pairs of subscripts of the vectors $(0 i)$ or $(\pi/2 i)$  by the single Greek subscripts $\a, \b$. 
\par

Since we have chosen the test vectors from the template bank, the scalar products appearing in Eq. (\ref{expect}) are just the ambiguity functions. Further these scalar products corresponding to the two phases are readily available from the output of the search pipeline. Accordingly, we define two ambiguity functions at the waveform vector locations as,
\be
\H_{0 \a} = (\h_0 (0), \h_{\a}),~~~~ \H_{\pi/2~ \a} = (\h_{\pi/2} (0), \h_{\a}) \,.
\label{ambfun}
\ee

We may then write the differences between the expected and observed correlations in terms of the ambiguity functions as follows:
\bea
\Delta c_{\a} (\x) &=&  c_{\a}^{o} (\x) - c^{e}_{\a} = (\x, \h_{\a}) - (\x, \h_0 (0)) \H_{0 \a} - (\x, \h_{\pi/2} (0)) \H_{\pi/2 ~\a} \,, \no \\
&\equiv& (\x, \Delta \h_{\a}) \,,
\label{delcamb}
\eea
where the projected vector $\Delta \h_{\a}$ now becomes,
\be
\Delta \h_{\a} = \h_{\a} - \h_0 (0) \H_{0 \a} - \h_{\pi/2} (0) \H_{\pi/2 ~\a} \,.
\label{dhalphaamb}
\ee
The covariance matrix can also be expressed in terms of the ambiguity functions. We obtain,
\be
C_{\a \b} = (\Delta \h_{\a}, \Delta \h_{\b}) = (\h_{\a}, \h_{\b}) - \H_{0 \a}\H_{0 \b} - \H_{\pi/2 ~\a}\H_{\pi/2 ~\b} \,.
\ee
We then just follow the steps identical to those followed for the generic $\chi^2$ and obtain the ambiguity $\chi^2$ as given by Eq. (\ref{final}). For convenience we also present the formula here as well,
\be
\chi^2 = \Delta c_{\a} [C^{-1}]^{\a \b} \Delta c_{\b} \,.
\label{amb}
\ee

Let us next estimate the computational costs involved in computing the ambiguity $\chi^2$. We note that all the scalar products with the data are available from the matched-filtering stage of the search pipeline. So no additional FFT cost is involved. In computing $\Delta c_{\a}$, we see from Eq.~(\ref{delcamb}) that ${\cal O}(p)$ operations are involved. The matrix $C$, the orthogonal matrix $O$ and the eigenvalues $\lambda_{\a}$ can be precomputed. This becomes easy in the spinless case where by choosing chirp-times as parameters, the scalar products weakly depend on the parameters. So we could divide the parameter space into a few regions and compute $C, ~O, ~\lambda_{\a}$ in each region. Then computation of $\Dc'$ and the multiplication of $\Dc$ by $O$ require ${\cal O}(p^2)$ operations, and finally computing $\chi^2$ requires ${\cal O}(p)$ more operations. Therefore, we have obtained a $\chi^2$ statistic with very little overhead costs. In the next section, we apply this statistic to actual numerical examples and examine how it works. 

\subsection{Ambiguity $\chi^2$ for the Newtonian waveform and single phase }
\label{Newton}

 For demonstrating how the ambiguity $\chi^2$ works, we consider the simplest possible case of the Newtonian waveform. The Newtonian waveform depends only on a single mass parameter, namely, the chirp mass $\M$ or, equivalently, the chirp time $\tau_0$, as defined in Eq.~(\ref{chrpt}). This reduction in one dimension of the parameter space is important because it allows us to exhibit the method conveniently via two-dimensional contour diagrams of ambiguity functions in the parameters $\tau_0$ and $t_c$. Secondly, we restrict ourselves to one phase (the dominant phase). There is no fundamental reason for doing this - we only do this for simplifying the procedure and demonstrating it in a lucid manner. Consider a trigger from a template with a certain $\vartheta$ and $t_c$ and assume that $c_0 > c_{\pi/2}$. If this is not the case then we can easily exchange the roles of the templates corresponding to $\phi_c = 0$ and $\phi_c = \pi/2$ in the expressions. Assuming, without loss of generality, that $c_0 > c_{\pi/2}$, we use the vectors $\h_{0 i}, ~i = 1, 2, ..., p$ corresponding to zero phase to construct the $\chi^2$ statistic. We drop the Greek subscripts for this subsection only.
\par

In Fig.~\ref{amb0pi} below, we have plotted the ambiguity functions $\H_0 (\Delta t_c, \Delta \tau_0)=  (\h_0 (0), \h_0 (t_c + \Delta t_c, \tau_0 + \Delta \tau_0))$ and $\H_{\pi/2} (\Delta t_c, \Delta \tau_0) =  (\h_{\pi/2} (0), \h_0 (t_c + \Delta t_c, \tau_0 + \Delta \tau_0))$ for the aLIGO PSD, where we have used the obvious notation of the subscript identifying the phase $\phi_c$. These ambiguity functions can be precomputed. 

\begin{figure}[ht!]
\begin{center}
\includegraphics[width=3.5in]{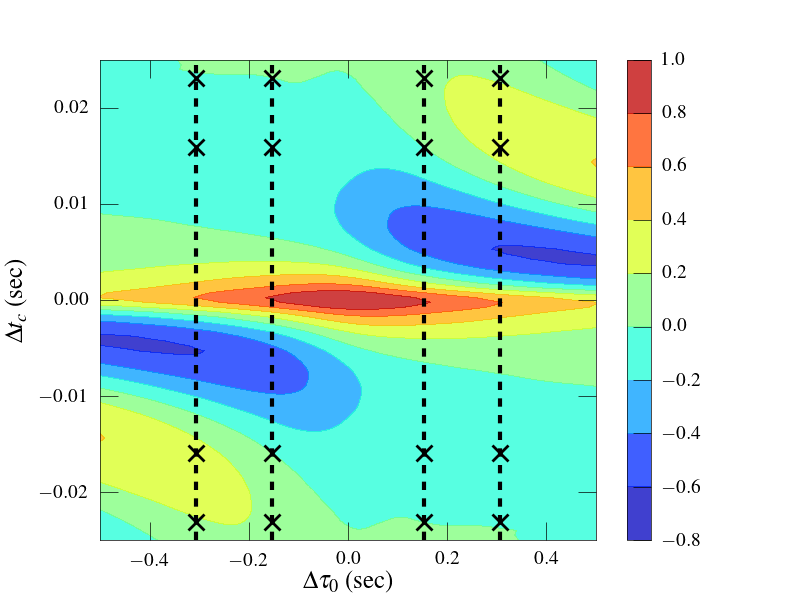}
\includegraphics[width=3.5in]{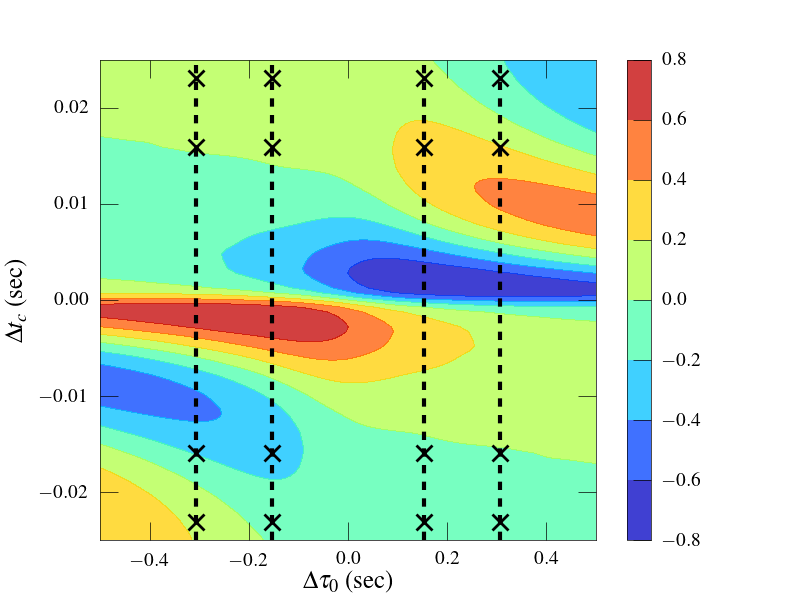}
\caption{Plots of the ambiguity functions $\H_0 (\Delta t_c, \Delta \tau_0)$ (left) and $\H_{\pi/2} (\Delta t_c, \Delta \tau_0)$ (right) around the trigger template for 1.4 - 1.4 $\Msun$ binary with $\Delta t_c = \Delta \tau_0 = 0$. For a mismatch of $\epsilon = 0.03$ and aLIGO noise, one finds $\Delta \tau_0 \sim 0.076$ sec. The templates in $\tau_0$ are placed at twice this distance. The dashed vertical lines show the points at which the filtered output is available. The crosses mark the templates chosen to construct the $\chi^2$ in the example given here. }
\label{amb0pi} 
\end{center}
\end{figure}

Let us consider the situation depicted in Fig.~\ref{amb0pi}. Let us consider a trigger signal with  amplitude $A = 10$ and dominated by $c_0$. We now need to choose waveform vectors around this trigger. Just to illustrate the workings of this test we choose only 4 points, i.e., $p = 4$. The filtered output is available at the points indicated by the dashed lines. Accordingly, we choose the points at $\Delta \tau_0 \simeq \pm 0.1534$ sec. This is twice the distance where the match falls to $0.97$. We also choose $\Delta t_c = \pm 0.0159$ sec. This gives 4 points placed symmetrically about the trigger in the form of a rectangle. We now calculate the expected values of $c_{0i}^{e}$ from the trigger correlations $c_0 (0), c_{\pi/2} (0)$ and the values of the ambiguity functions at these points. We then read off observed values $c_{0i}^{o}$ from the filtered output and take differences to obtain the vector $\Dc$. The covariance matrix computed for these points is:
\be
C = \left[ \begin{array}{cccc} 0.9243 & 0.0138 & 0.5781 & -0.0161 \\ 0.0138 & 0.9972 & 0.00046 & 0.5781 \\ 0.5781 & 0.00046 & 0.9972  & 0.0138  \\ -0.0161 & 0.5781 & 0.0138 & 0.9243 \end{array} \right] \,.
\ee  
In order to deduce $\Dc'$ we require the orthogonal transformation $O$ that diagonalizes $C$. This turns out to be:
\be
 O = \left[ \begin{array}{cccc} -0.5185 & 0.5123 &  0.4808 & -0.4874 \\ 
  0.4808 &  0.4874 &  0.5185 & 0.5123 \\
  0.4808 & -0.4874 &  0.5185 & -0.5123 \\ 
 -0.5185 & -0.5123 & 0.4808 & 0.4874 \end{array} \right] \,.
\ee
As explained before, this transformation is the key in extracting the eigenvalues of $C$, which in this case are $0.35941223, ~0.40363689, ~1.53366376, ~1.5465145$, in increasing order. 
\par
We can now use these results to compute the $\chi^2$ for any data vector $\x$. One can verify that when $\x$ is just the signal, one obtains $\chi^2 = 0$. For a realization of Gaussian stationary noise with mean zero and aLIGO PSD, we found the average value to be $\langle \chi^2 \rangle \sim 4.27$. The expected value is 4 because we have 4 degrees of freedom. The value we obtained in this case is close to the expected value. For a  sine-Gaussian glitch with the quality factor $Q = 5$ and central frequency $f_0 = 60$~Hz that gives a trigger SNR of 10, the $\chi^2 \sim 187$. 
\par
\begin{figure}[ht!]
\begin{center}
\includegraphics[width=3.5in]{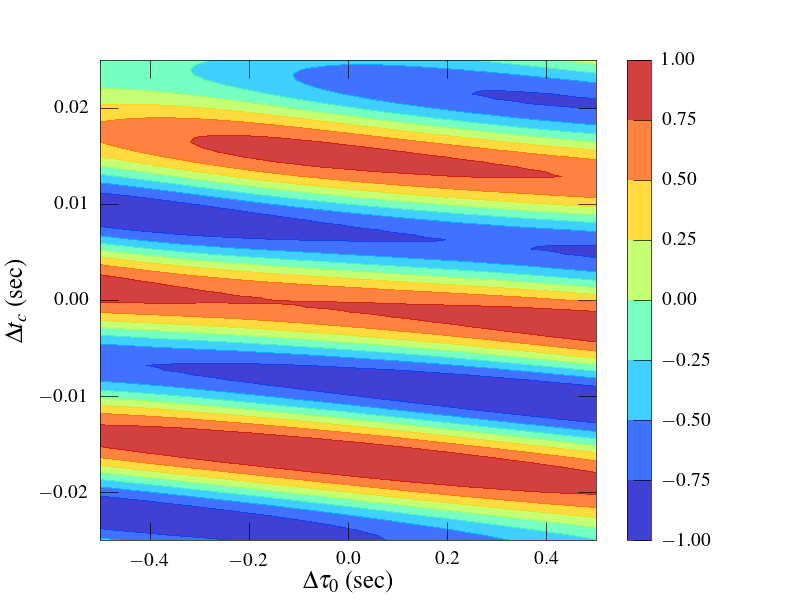}
\caption{The figure shows the match of the sine-Gaussian glitch with $Q = 5$ and $f_0 = 60$ Hz with the normalized phase zero waveforms. The match is seen to be very different from that of a signal indicating that the ambiguity 
$\chi^2$ is effective in separating glitches from signals.}
\label{glitch} 
\end{center}
\end{figure}

We can increase the number of waveform vectors to a larger number, say $p = 16$. We then choose the points $\Delta \tau_0 = \pm 0.1534, \pm 0.3068$ and $\Delta t_c = \pm 0.01589, \pm 0.02315$. These 16 points are also symmetrically placed around the trigger. While we do not explicitly present the covariance matrix, the orthogonal transformation matrix, and the eigenvalues, since that is cumbersome, nonetheless we evaluate the $\chi^2$ for the aforementioned sine-Gaussian glitch with these waveform vectors and find its value to be $\sim 365$, which is much greater than the expected value of 16. These large values of $\chi^2$ arise because the match of the glitch with the vectors surrounding the trigger is very much different from the ambiguity functions. We show this match for $\phi_c = 0$ surrounding the trigger in Fig. \ref{glitch}.

\subsection{Effect of mismatch}
\label{ambmsmtch}

 In this subsection we investigate the effect of mismatch between the signal and the template parameters on the ambiguity
$\chi^2$. A mismatch can occur because, in principle, the signal can have any parameters continuously distributed over the parameter space, while the templates are necessarily placed at discrete set of points stipulated by the minimal match~\cite{Owen}. However, owing to the mismatch, in our $\chi^2$ construct, the signal will not be subtracted completely, and there will be a residual which we have denoted by $\delta \h$ in Sec.~\ref{msmtch}. This is expected since the signal will not be orthogonal to the chosen subspace. 
\par

\begin{figure}[ht!]
\begin{center}
\includegraphics[width=3.5in]{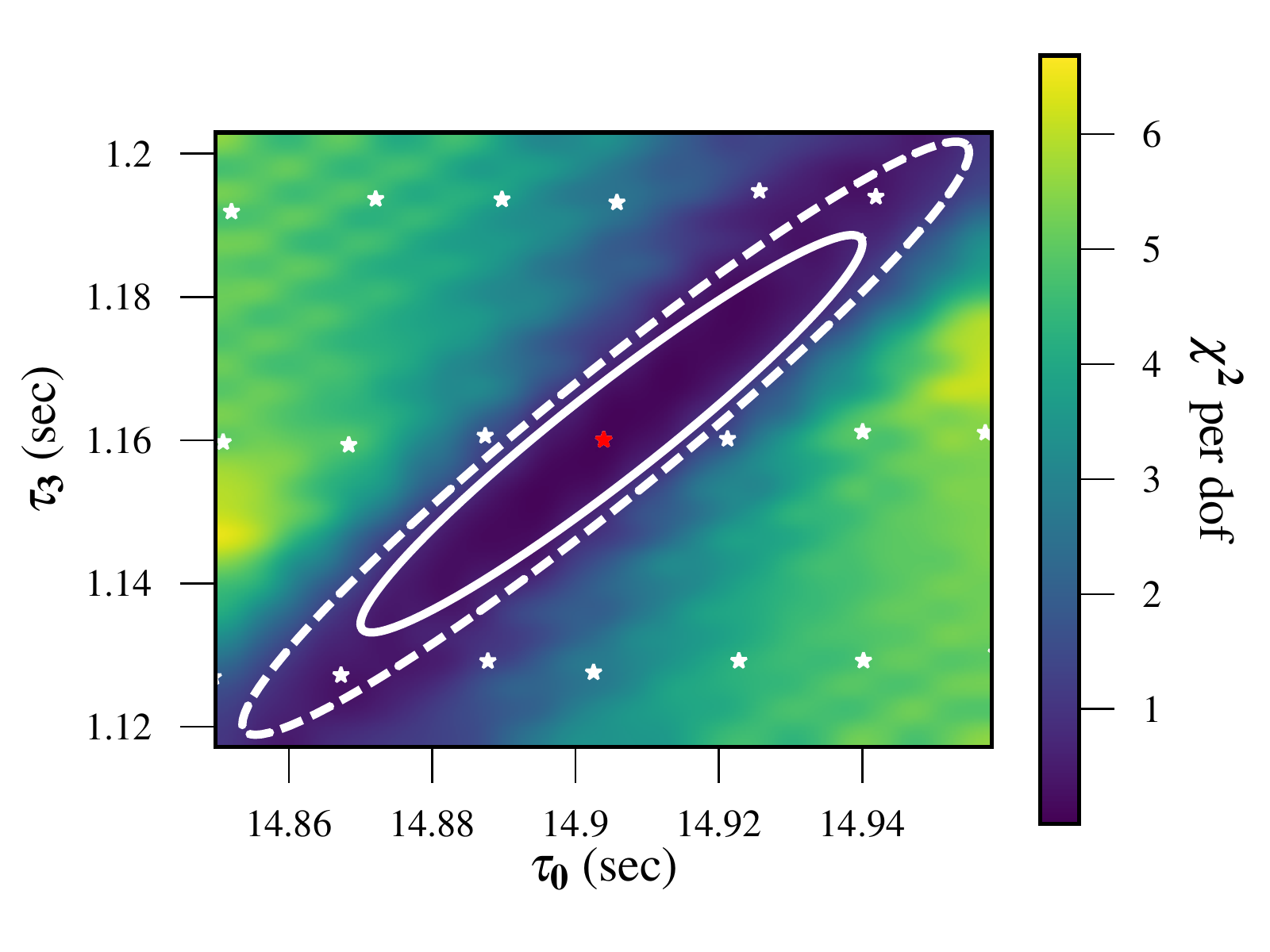}
\includegraphics[width=3.4in]{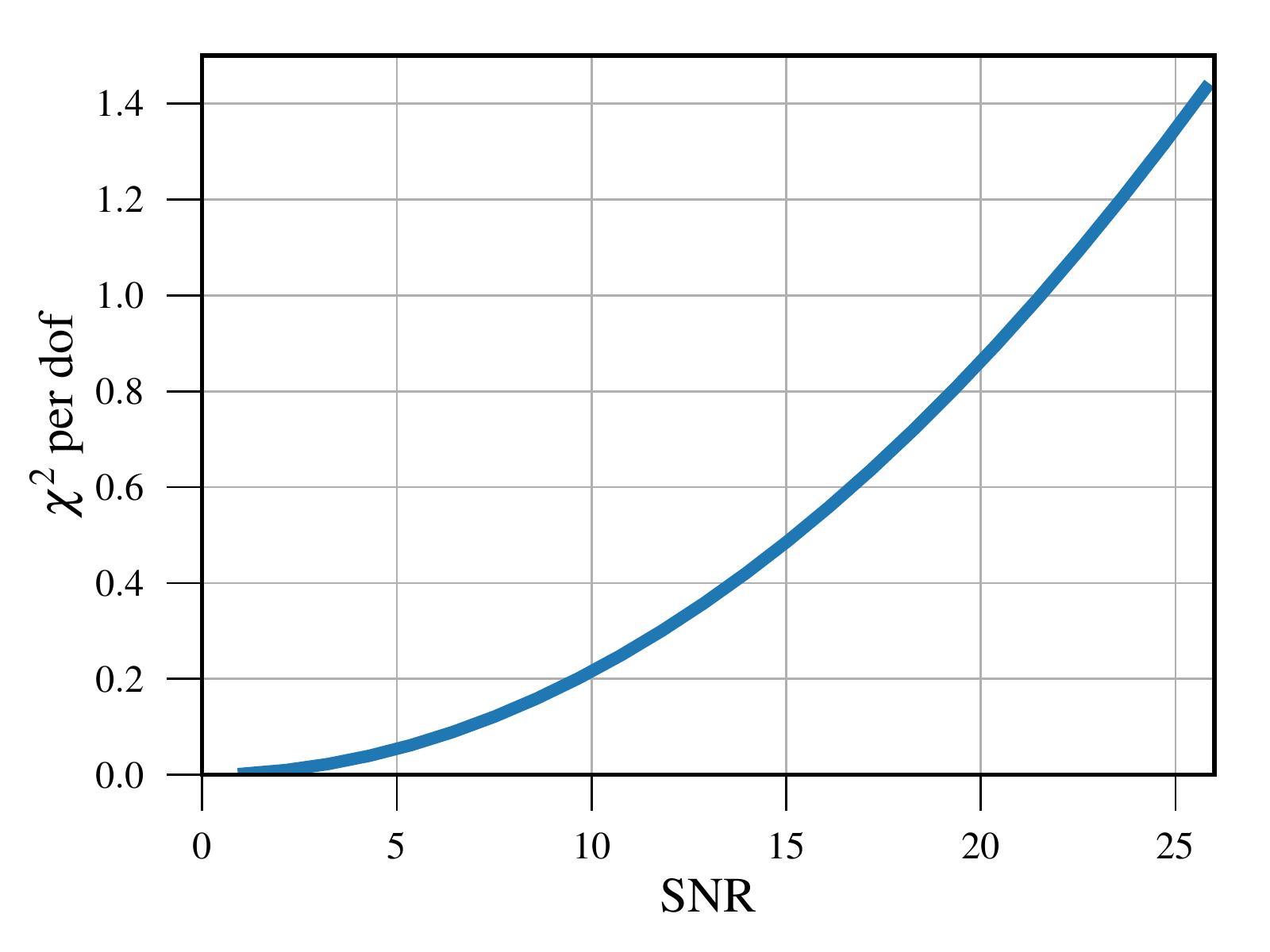}
\caption{The left figure shows the effect of mismatch on the ambiguity $\chi^2$ (in colour) as a function of the parameters $\tau_0$ and $\tau_3$ in the absence of noise. The trigger template is placed in the center of the plot and shown as a red star. For reference, ellipses at matches of 0.97 (inner) and 0.95 (outer) are shown. The 6 adjacent templates (shown as white stars) lie in the blue region indicating that the effect of mismatch is sufficiently small ($\chi^2 \sim 1$ per degree of freedom) for the $\chi^2$ to be effective. The right figure shows the $\chi^2$ per degree of freedom versus the SNR for a fixed mismatch of 0.02. We observe that the $\chi^2$ value increases quadratically with SNR.}
\label{mismatch} 
\end{center}
\end{figure}

The results of our investigation into the effects of mistmatch are presented in Fig.~\ref{mismatch}, where 
we use the spinless TaylorF2 approximant to model both signal and template waveforms. At any given instance, we take the signal to have parameters corresponding to a single point in the left plot of that figure. However, we always assume that the trigger template has parameter values shown by the red star in that plot. We numerically compute the ambiguity
$\chi^2$ values, depicted in colour, in the vicinity of the trigger template for signals with parameters chosen in the ranges shown in the plot.
We assume noise to be zero. Note that we have chosen the trigger template well inside the allowed space parameterized by the two chirp times $\tau_0$ and $\tau_3$. 
The templates shown in white, and surrounding the trigger template, are search templates, which may or may not be the same templates that are used to calculate the ambiguity $\chi^{2}$ statistic. The signal can be anywhere in the neighbourhood of the trigger template. Here, for the purpose of illustration, we use 14 of these search templates to compute the ambiguity $\chi^2$, in aLIGO noise PSD and with a lower cut-off frequency of 30 Hz; however, there are only 8 degrees of freedom because after diagonalising the covariance matrix we consider only those eigenvalues that are larger than $0.05$.\footnote{We use this threshold on the eigenvalues in all of our numerical simulations in order to select the degrees of freedom.} 
Ellipses for 0.97 (inner) and 0.95 (outer) match with the trigger template are also shown for reference. The templates that have the best chance of being triggered by a signal with the same parameters as the trigger template lie along or close to the semi-major axis of the match ellipses. One immediately observes that if a signal had the same parameter values as any of the six templates surrounding the trigger template (centre) and lying in the blue region, then its $\chi^2$ per degree of freedom would be $\sim 1$. One concludes from this observation that a small mismatch of this order does not pose any serious problem for implementing the ambiguity $\chi^2$.    

We have also investigated how the $\chi^2$ depends on the SNR in right panel of Fig.~\ref{mismatch}. We take a fixed match of 0.98 for this purpose and vary the SNR. We find that the $\chi^2$ increases quadratically with the SNR in accordance with Eq. (\ref{bound}). 

\subsection{$\chi^2$ analysis for simulated data}

In this section we test the performance of the ambiguity $\chi^2$ on simulated data. We prepare the simulated data segments with Gaussian colored noise using aLIGO design PSD. The lower cut-off is set to be 30 Hz, which is same as that chosen for first observing run for the aLIGO detectors. In the so constructed Gaussian noise data we inject simulated GW signals and glitches. We have used non-spinning templates for our simulations. For the binary black hole (BBH) case, we choose component masses uniformly sampled in the range $5 ~\Msun - 10~ \Msun$. For the neutron star -  black hole (NSBH) case, the black hole masses are uniform in $5 ~\Msun - 10~ \Msun$ while neutron star masses are uniformly sampled in the range $1.3 ~\Msun - 2~ \Msun$. There are 60 BBH signals and 40 NSBH signals injected in the simulated noise whose SNRs are ranging between $4 ~-~ 40$ assuming an uniform volume distribution of sources. Further, we inject  $100$ Sine-Gaussian (SG) glitches~\cite{CBDL,BDGL,Powell15a} for both BBH and NSBH cases. One of the reasons for our choice of SG glitches is that the sine-Gaussians form a useful basis on which many of the glitches in real data have strong projections~\cite{Chatterji:2004qg,Chatterji:2005thesis}. The central frequencies for these SG injections are chosen in the range $[30, 500]$ Hz, spaced uniformly in logarithmic scale while the quality factors are uniform in $[5, 10]$. We also inject 100 Gaussian glitches~\cite{Powell15a} in both cases with standard deviations ranging from $20$ to $100$. For both the glitch injections, the SNR is maximized over the template banks such that it ranges from $4$ to $40$ as is done for GW injections. While such glitch injections are useful to an extent, real data must eventually be used for a proper test of the performance of these discriminators.

\par
We also study the performance of the ambiguity $\chi^2$ when the data contain only colored Gaussian noise. For this purpose we take $100$ independent realizations. We compute the ambiguity $\chi^2$ for these injections as follows: In the ambiguity $\chi^2$ (or more generally in a bank $\chi^2$), the trigger template is a random variable. However, when there is a trigger with SNR above a threshold of 7 or 8, the distribution for
the trigger template is highly peaked and the trigger template is essentially pinned down. This
happens for both GW signals and glitches. Therefore, the ambiguity $\chi^2$ has approximately $\chi^2$ 
 distribution for signals, non-central $\chi^2$ distribution for glitches with non-centrality parameter $\| \g_\S \|^2$. However, in the case of Gaussian noise where the SNRs are low,
the trigger template jumps from one noise realization to other such that the subspace $\S$ changes violently from one noise realization to other. Thus, in the $\chi^2$ the terms of the form $\n \cdot \e_\alpha$ are not Gaussian distributed at all as
$\e_\alpha$ is a random variable. Consequently, the $\chi^2$ possibly has some other distribution (which so far we have not found) giving rise to high values of the statistic. Note that in the traditional $\chi^2$ the subspace $\S$ is fixed before the search such that
Gaussian noise gives a true $\chi^2$ distribution. We therefore arbitrarily fix the trigger template. This gives a true $\chi^2$ distribution for Gaussian noise injections.
\par

\begin{figure}[H]
\begin{center}
\includegraphics[width=3.5in]{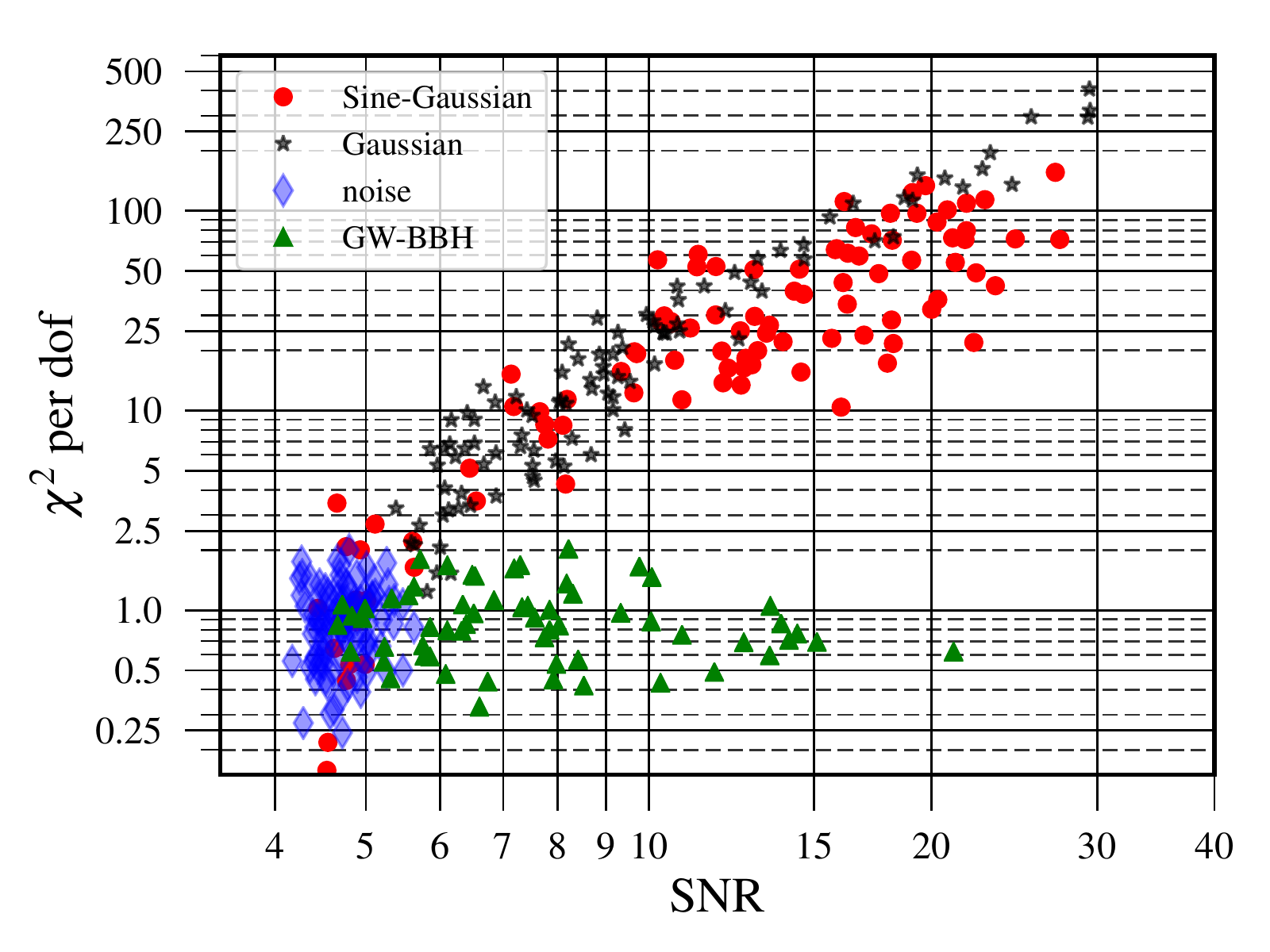}
\includegraphics[width=3.5in]{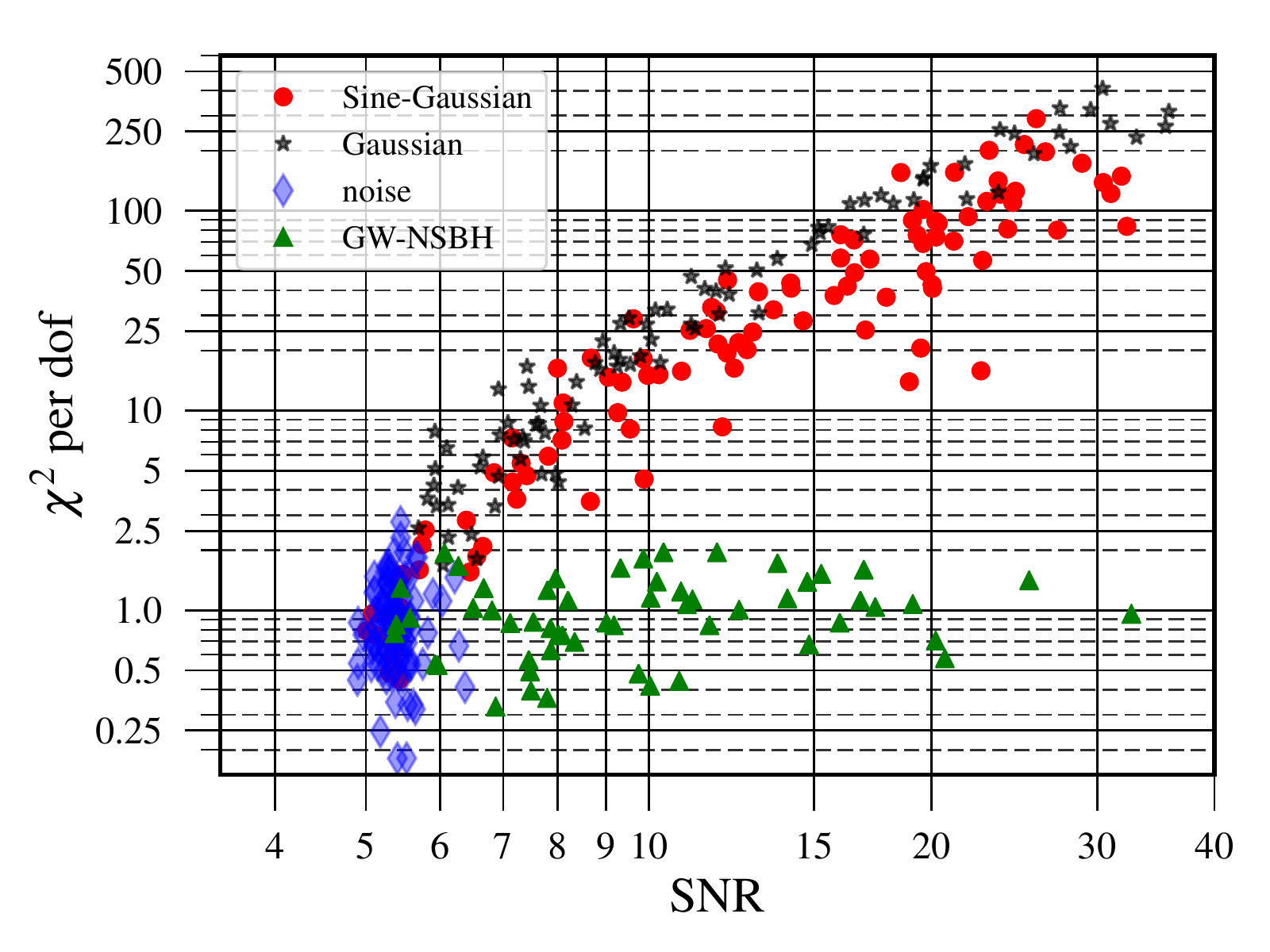}
\caption{Figures showing the $\chi^2$ per degree of freedom versus SNR for triggers of CBC signals (green triangles), sine-Gaussian glitches (red circles), Gaussian glitches (grey stars) and Gaussian noise (blue diamonds). The left figure pertains to BBH injections and the right one to NSBH injections. We observe that in general the ambiguity $\chi^2$ separates out CBC signal triggers from noise triggers.}
\label{scatter} 
\end{center}
\end{figure}

In the Fig.~\ref{scatter}, we plot ambiguity $\chi^2$ per dof as a function of SNR for BBH case (left) and NSBH case (right). The green triangles are triggers from the CBC signals and we see that they generally have low $\chi^2$ value. The red circles are the triggers from the sine-Gaussian glitches and are seen to have high $\chi^2$. The triggers from Gaussian noise are shown as blue diamonds which lie in the neighborhood of the origin - they have both low SNR and $\chi^2$. We also plot triggers from Gaussian glitches which also have high $\chi^2$ per dof, as expected. 
Therefore, we observe that the ambiguity $\chi^2$ has the important property of separating the noise triggers from the signal triggers and thus is a good candidate as a $\chi^2$ discriminator. For better performance it may be possible to tune the ambiguity $\chi^2$.

Note that fixing the trigger template when the data are purely Gaussian noise was done only to show that we get a $\chi^2$ distribution in such a case. If we do not fix the trigger template as remarked before, we only get a higher value for the $\chi^2$ statistic, but since it is in fact noise, it will not affect the implementation of the ambiguity $\chi^2$. 

\section{Generalization to the coherent multi-detector case}
\label{multidetector}

It is easy to generalize the framework to the coherent multi-detector case~\cite{Bose:1999pj,PDB01}.
We use the results from the above works to generalize to the multi-detector case. We do not however go into the details. Following \cite{Bose:2011km,HF11} we can write the signal in detector $I$ as,
\be
h^I (t) = \A^\mu h_\mu^I (t) \,, 
\ee
where $\A^\mu$ are amplitudes depending on the distance to the source, initial phase, polarization angle and inclination of the source and $h_\mu^I (t)$ are defined by,
\be
h_1^I (t) = F^I_+ h_0 (t^I), ~h_2^I (t) = F^I_{\times} h_0 (t^I), ~h_3^I (t) = F^I_+ h_{\pi/2} (t^I), ~h_4^I (t) = F^I_{\times} h_{\pi/2} (t^I) \,,
\label{basis}
\ee
where $F^I_{+, \times}$ are the antenna pattern functions of detector $I$ and $t^I$ is the retarded time in detector $I$. 
\par
Consider that there are $M$ detectors, then the multi-detector signal vector can be written as,
\be
\h (t) = (h^1 (t), h^2 (t), ..., h^M (t)) \,,
\ee
where now the signal vector belongs to the direct sum of Hilbert spaces $\D_I$, where $\D_I$ is the Hilbert space of the data trains pertaining to the detector $I$. We denote this space by,
\be
\Dn = \D_1 \oplus \D_2 \oplus ... \oplus \D_M \,.
\ee
The vector space $\Dn$ becomes a Hilbert space if we define a scalar product on it. For uncorrelated noise between different detectors, a  natural scalar product on $\Dn$ is just the sum of the scalar products corresponding to  individual detectors \cite{HF11,PDB01}. This is the scalar product we use on $\Dn$ in the context of the network.
\par

From Eq. (\ref{basis}), the 4 network signal waveform vectors $\h_\mu \in \Dn$ are given by,
\be
\h_\mu = (h^1_\mu (t), h^2_\mu (t), ..., h^M_\mu (t)) \,, ~~~~\mu = 1, 2, 3, 4 \,.
\ee
As shown in \cite{HF11}, in the dominant polarization frame (which is also done in \cite{PDB01}), these four vectors are mutually orthogonal (though not necessarily orthonormal). This construct then leads to the coherent statistic which is also essentially the $F$ statistic (except for a factor of 2) first defined for continuous wave sources in Ref.~\cite{JKS98}.   
\par

To construct the $\chi^2$ for a network, it is only necessary to select a subspace $\S_{\rm network} \subset \Dn$ which is orthogonal to each of the $\h_{\mu}$. This can be easily done. Then the network $\chi^2$ is just the square of the $L_2$ norm of a data vector $\x \in \Dn$ projected into the subspace $\S_{\rm network}$. 
\par

In the single detector case we had just two basis waveforms corresponding to the phases $\phi_c = 0,~\pi/2$. In the coherent multi-detector case we have four basis waveforms corresponding to the two phases and the two polarizations $`+',~`\times'$. If so desired, an ambiguity $\chi^2$ for the multi-detector case can be designed based on selecting only one of the signal waveforms $\h_\mu$ as was done in the single detector case in subsection  \ref{Newton}. We do not pursue this point here any further and leave it for future investigation. 

\section{Effect of small mismatch in parameters between the signal and the template}
\label{msmtch}

Although the kinematical parameters $t_c$ and $\phi_c$ can be searched over continuously by the FFT and analytical maximization, the mass parameters are searched with a discrete bank of templates. In practice, the templates are laid such that the match maximized over $t_c$ and $\phi_c$ is at least $97\%$ or the mismatch is at most $3\%$. It is very unlikely that a signal will have exactly the parameters as those of the templates in the bank - in general there will be a mismatch. In this subsection, we estimate the effect of this mismatch on the value of the $\chi^2$. Although the $\chi^2$ vanishes for a signal matching the template - the subspace $\S$ is orthogonal to the template - it will not in general be orthogonal to the signal because of the mismatch. Nevertheless, if the mismatch is small, we will show that this effect on $\chi^2$ is small and we will derive a bound on this value. 
\par

We perform the analysis in general. We consider a signal $\s$ with parameters $\vartheta^a + \Delta \vartheta^a$ and  amplitude $A$, then we have $\s (\vartheta^a + \Delta \vartheta^a) = A \h (\vartheta^a + \Delta \vartheta^a)$, where $\h (\vartheta^a)$ denotes a normalized waveform.  The trigger template has parameters $\vartheta^a$ and we write $\h (\vartheta^a + \Delta \vartheta^a) = \h (\vartheta^a) + \delta \h$. Since both $\h (\vartheta^a + \Delta \vartheta^a)$ and $\h (\vartheta^a)$ are normalized, to the first order, $\delta \h$ is orthogonal to $\h (\vartheta^a)$. {\it However, it is important to realize that $\delta \h$ may not lie in the subspace $\S$ corresponding to the trigger template.} In fact to the first order, $\delta \h \in \Nchi$. We will have more to say about this later in this subsection. For now, we relate the norm of $\delta \h$ to the ambiguity function. We have the following relation:
\bea
\| \delta \h \|^2 &=& \| \h (\vartheta^a + \Delta \vartheta^a) - \h (\vartheta^a) \|^2 \,, \no \\
                  &=& \| \h (\vartheta^a + \Delta \vartheta^a) \|^2 + \| \h (\vartheta^a) \|^2 - 2 (\h (\vartheta^a), \h (\vartheta^a + \Delta \vartheta^a)) \,, \no \\
                  &\equiv& 2 (1 - \H (\vartheta^a, \Delta \vartheta^a)) \,,
\eea 
where $\H (\vartheta^a, \Delta \vartheta^a) = (\h (\vartheta^a), \h (\vartheta^a + \Delta \vartheta^a))$ is the ambiguity function. The mismatch $\eps$ is related to the ambiguity function by:
\be
\H (\vartheta^a, \Delta \vartheta^a) \geq 1 - \eps \,.
\ee
Therefore, we have $1 - \H (\vartheta^a, \Delta \vartheta^a) \leq \eps$ which leads to the following inequality:
\be
\| \delta \h \|^2 ~\leq ~ 2 \eps \,.
\ee
From the above relation, we can now obtain a bound for $\chi^2$. The $\chi^2$ of the signal $\s (\vartheta + \Delta \vartheta^a)$ mismatching with the template $\h (\vartheta^a)$ is,  
\be
\chi^2 (\s) = \chi^2 (A (\h + \delta \h)) \,.
\label{loose}
\ee
Projecting orthogonal to $\h$, we must have:
\be
\chi^2 (\s) = A^2 ~ \| {\delta \h}_\S \|^2 \leq A^2~ \| \delta \h \|^2 \leq 2 A^2 \eps \,.
\label{bound}
\ee
This bound has been obtained in Ref.~\cite{Allen} (Eq. (6.24) therein) for $\chi^2_t$. But here we have obtained this bound generally without referring to any specific $\chi^2$. Even with this bound, one can see that for $\eps = 0.03$ and $A = 10$, $\chi^2 (\delta \h) \leq 6$. For the ambiguity $\chi^2$ proposed here with 16 templates, a numerical evaluation leads to $\chi^2 \sim 4.3$. 
\par

The above result does not depend on the degrees of freedom because we have obtained this bound from 
$\delta \h$  which is in $\Nchi (\h)$. However, in Ref.~\cite{Allen} a tighter bound has been obtained. In general we may be able to obtain a better bound by projecting onto the subspace $\S$ since $\| \h_{\S} \|^2 < \| \h \|^2$. But now the bound could depend on $p$, the dimension of $\S$. This is what happens as is seen from the results obtained in Ref.~\cite{Allen}. Here we sketch the arguments given in Ref.~\cite{Allen} in the context of our framework.  
\par

For brevity, let us write the normalized mismatched signal vector as $\h' = \h + \delta \h$. We now project 
$\h'$ onto the $p - 1$ dimensional subspace $\S$ and obtain $\h'_{\S}$, then $\chi^2 = A^2 \| \h'_{\S} \|^2$. In order to make connection with Ref.~\cite{Allen} we need to go to $p$ dimensions. We can write the basis vectors $\e_\a, ~\a = 1,~2, ... , p$, as linear combinations of $\Delta \h_\b$ in terms of an orthogonal $p \times p$ matrix $O$ given by,
\be
O = \left[ \begin{array}{cccccc} \frac{1}{\sqrt{2}} & - \frac{1}{\sqrt{2}}  &  0 & 0 & ... & 0 \\ 
 \frac{1}{\sqrt{6}} & \frac{1}{\sqrt{6}} & - \frac{2}{\sqrt{6}} & 0 & ... & 0 \\
  ... & ... & ... & ...& ...& ...  \\ 
 \frac{1}{\sqrt{p}} & \frac{1}{\sqrt{p}} & ... & ... & ... & \frac{1}{\sqrt{p}} \end{array} \right]
\ee
That is,
\be
[\e_1, \e_2, ..., \e_{p - 1}, \e_p ]^T = \sqrt{p}~O~ [\Delta \h_1, \Delta \h_2, ..., \Delta \h_p ]^T \,,
\ee
where the symbol $T$ denotes the transpose; it transposes the row vectors to column vectors and also effects $O^T O = 1$. Note that the last vector $\e_p = 0$ because of Eq. (\ref{bins}). Following the notation in Ref.~\cite{Allen} one finds that,
\be
\h' \cdot \Delta \h_\b = \h' \cdot \left(\h_\b - \frac{\h}{p}\right) = \left(\vartheta_\b - \frac{1}{p}\right) \cos \theta \equiv \omega_\b \cos \theta \,,
\ee 
where $\cos \theta = 1 - \epsilon$ is the match, and $\theta$ is the ``angle" between $\h$ and $\h'$, i.e., $\h \cdot \h' = \cos \theta$. Writing $\Bom = [\omega_1, \omega_2, ... , \omega_p]^T$ and $\h'_{\S}$ as column vectors we easily get,
\be
\h'_{\S} = [\h' \cdot \e_1, \h' \cdot \e_2, ... , \h' \cdot \e_p ]^T = \sqrt{p} ~O~ \Bom ~ \cos \theta \,.
\ee
The contribution of the mismatch to $\chi^2_t$ is,
\be
\chi^2_t = A^2 ~\| \h'_{\S} \|^2 = A^2~ {\h'}_{\S}^T ~\h'_{\S} = p~A^2 ~\Bom^T~ O^T~O~\Bom~ \cos^2 \theta = p~A^2~ \cos^2 \theta~ \Bom^T \Bom = p~A^2 \cos^2 \theta ~\sum_{\b = 1}^p \omega_\b^2 \,.
\ee
In Ref.~\cite{Allen}, the constraints on $\omega_\b$ are derived by assuming that both the signal and template contribute the same power to each frequency bin. To the lowest order in $\epsilon$, they are the following:
\be
- \frac{\epsilon}{p} ~~ \leq ~~ \omega_\b ~~ \leq ~~ \frac{2}{p},~~~~~~~~~ \sum_{\b = 1}^{p} \omega_\b = 0 \,.
\ee
Subject to these constraints the $\chi^2_t$ can be maximized. This happens when $\omega_1 =  \epsilon (p -1)/p,~ \omega_2 = \omega_3 = ... = \omega_p = - \epsilon/p$ and the result to the lowest order in $\epsilon$ is  
$\chi^2_t \sim (p - 1) \epsilon^2 A^2$. (The $\cos^2 \theta$ factor drops out because we are interested in the lowest order in $\epsilon$.) This bound is smaller than the general bound $2 \epsilon A^2$ only when $p \epsilon < 2$, that is, the degrees of freedom $p$ are not too large. This is the second bound obtained for $\chi^2_t$. We do not know whether some analogous bound exists in general.
\par

We could go even further. Let us first pick a waveform $\h_0 \in \P$. If we were to choose the subspace ${\S}_0$ in such a way that it is orthogonal  to {\it both} $\h_0$ and $\h_0 + \delta \h$ (this could be easily achieved since there is a lot of freedom in the choice of $\S$), then the effect of the mismatch could be completely nullified. However, there is not one $\delta \h$, but several waveforms $\h$ centered at say $\h_0$, with 
$\delta \h = \h - \h_0$, which lie within the region of mismatch determined by $\eps$. This region is given by:
\be
E_0 = \{\h \in \P/ ~\| \h - \h_0 \|^2 < 2 \eps \} \,.
\ee
We denote it by $E_0$ since it is essentially shaped like an ellipsoid (or hyper-ellipsoid in higher dimensions) for small mismatches, because expanding the ambiguity function to the quadratic order is sufficient. Note that $\| \delta \h \|^2$ is a quadratic form in $\Delta \vartheta^a$ and is in fact $2 \times g_{a b} \Delta \vartheta^a \Delta \vartheta^b$, where $g_{a b}$ is the metric. (The factor of $2$ comes because the metric is defined with this factor in Ref.~\cite{Owen}. However, if one uses the definition of the metric from Ref.~\cite{BSD}, this factor of $2$ does not appear). Topologically, $E_0$ is an open sphere with radius $\sqrt{2 \eps}$.  The value of $\chi^2$ on $E_0$ will not be identically zero but could be made close to zero, by choosing the subspace ${\S}_0$ at $\h_0$ judiciously mitigating the effect of the mismatch. In the maximization procedure of the statistic over the kinematical parameters, one may in fact need to focus only on a subset of $E_0$ so that the problem could be less involved.
\par
We remark that the mismatch parameter $\eps$ used here is quadratic in the amplitude just as in Ref.~\cite{Allen} and, thus, appears linearly in the bound in Eq.~(\ref{loose}). However, in Ref.~\cite{HF11}, the mismatch parameter $\eps$ is used in the amplitude and, therefore, appears as $\eps^2$ in the bounds on the $\chi^2$. It is also noteworthy that there may be reasons for mismatch other than the discreteness of the template bank: For instance, the model of the waveform may differ from the actual astrophysical signal, in which case a systematic error would be introduced, thereby, leading to higher values of $\chi^2$. We have not addressed this case here.

\section{Conclusions}
\label{sec:conclusions}
 
  We have presented for the first time a general framework for unifying $\chi^2$ discriminators that also unravels how new ones can be generated straightforwardly. We showed that these discriminators have the underlying mathematical structure of a vector bundle. The $\chi^2$ statistic is just the $L_2$ norm of the data vector projected onto a subspace of the Hilbert space of data vectors orthogonal to the trigger template. This underscores the fact that the basis of this subspace is unimportant. Since the trigger could occur anywhere in the parameter space, we are dealing with a collection of subspaces orthogonal to each vector tracing out the signal manifold - a smooth choice leads to a vector bundle. The $\chi^2$ can be visualized as a non-negative real valued function on sections of the fibre bundle, which are in fact the vector fields obtained by the projection of the data vectors onto the subspaces.  Apart from the elegance, the important practical insight that emerges from this formulation is the enormous flexibility available in defining $\chi^2$ discriminators. We expect that this freedom could be used in {\it tuning} the $\chi^2$ so that it can discriminate more decisively against frequently occurring glitches or glitches which pose difficulties to the search algorithms. In this context, we would like to point out that the work which relates template bank triggers to the sine-Gaussian glitch parameters \cite{BDGL,CBDL}. This may turn out to be useful in tuning the $\chi^2$ to sine-Gaussian glitches, because it relates regions of signal parameter space which respond to the parameters of the glitches. The problem is of course more general: since we have a family of $\chi^2$ to choose from, the tuning of the $\chi^2$ to real data is intimately linked. One needs to optimize the $\chi^2$ to those which are most effective on real data. This also raises a question: why the traditional $\chi^2_t$ performs so well. We believe our general framework would be useful in understanding this issue. Our near future goal is to make headway in these general directions.   
\par

Using this formulation, we have constructed a generic $\chi^2$ discriminator by choosing an arbitrary set of vectors in $\D$ and then projecting out their components parallel to the trigger templates. For instance we could choose these vectors in the direction of the glitches - say sine-Gaussians - and by subtracting out the projections on the trigger template construct $\S$ orthogonal to the trigger template. We could in fact choose the vectors to be a sine-Gaussian basis. This would be an useful $\chi^2$ to discriminate against sine-Gaussians. Such a $\chi^2$ would be viable if the cost in computing the $\chi^2$ is small compared with the cost of the CBC search.  Further, we have also defined a performance measure for a generic $\chi^2$ - a conditional probability - if the glitches can be modeled. We have obtained this probability explicitly for the sine-Gaussian glitches.
\par

By choosing the vectors in $\D$ from the template bank, and carrying out the construction as in the generic case, we have proposed a new $\chi^2$ - the ambiguity $\chi^2$ - based on the behavior of the ambiguity function. This is because the matched filter outputs of a bank of templates centered around the parameters of a strong enough signal sample the ambiguity function of that signal. If a trigger corresponds to a glitch and not a signal then those outputs will not, in general, follow the ambiguity function of a signal with the parameters of the triggered templates. The advantage of ambiguity $\chi^2$ is that its computation entails negligible overhead cost since most of the inputs required to compute it are available from other steps of the search~\cite{Allen:2005fk,Usman:2015kfa}. 
\par
 
We have also investigated the effect of the mismatch between the signal and the templates on the ambiguity $\chi^2$. The reason of mismatch is generic because in the intrinsic parameters like masses or spins, the parameter space is covered with a discrete bank of templates, and therefore, the signal parameters are unlikely to match with any particular template. This results in a higher value of $\chi^2$. We have shown that the effect of the mismatch does not pose any serious problem to the implementation, if the templates lie sufficiently close - say within the usual mismatch of $0.03$. The $\chi^2$ value per degree of freedom is at most of the order of unity in the noise free case, if the signal parameters mismatch up to the adjacent template. In fact, although we have not optimized for this, another advantage of the freedom in choice of the ambiguity $\chi^2$ is that we may be able to mitigate the effects of mismatch of the signal with the templates
\par

To demonstrate the power of ambiguity $\chi^2$ we carried out an extensive set of simulations with CBC signals and mock glitches separately added to data that was otherwise statistically Gaussian and stationary. The CBC signals were simulated with a wide range of astrophysical parameters corresponding to BBH and NSBH systems, involving stellar-mass black holes. The glitches were simulated to be of the Gaussian and sine-Gaussian types, with parameters (such as quality factor or central frequency) observed in real glitches. These data were then matched filtered with NSBH and BBH template banks, as in real searches. Results presented here show that the glitches are clearly separated from the signals. This is true even when the signals do not match the template parameters. Also the triggers from Gaussian noise lie in the neighborhood of the origin. Thus, in this simulated numerical experiment, the ambiguity $\chi^2$ is effective and yields encouraging results. Although our numerical simulations have been performed for the spinless TaylorF2 waveform, our discriminator is applicable to a wider class of waveforms - it has been successfully tried out for other spinless waveforms, for example, IMRPhenomD \cite{PhnmD}. We expect that the ambiguity $\chi^2$ should be applicable more generally to spinning waveforms. 
\par

 We also point out how the ambiguity $\chi^2$ has advantage over the bank $\chi^2$ \cite{Hanna} and the autocorrelation $\chi^2$ \cite{GSTLAL}. In the bank $\chi^2$, in order to have the test templates as mutually orthogonal as possible, the templates must be well spread out in the parameter space. But then several of the templates tend to be away from the trigger (almost zero match with the trigger template) but these may also give zero match with the noise transient as well, reducing the discriminatory power of the $\chi^2$; these degrees of freedom of the $\chi^2$ do not contribute to the $\chi^2$ value. On the other hand, for the ambiguity $\chi^2$ we have the freedom to choose templates near the trigger template, not worrying about orthogonality. Therefore, we believe that it would be more  effective in discriminating signals against the noise transient - at least the signal will have non-zero projection on the selected template vectors. It is clear that the directions probed by the $\chi^2$ should either be those of the signal or the glitch in order that the $\chi^2$ be effective. Moreover, if the test templates are not chosen to be sufficiently orthogonal, the effective number of degrees of freedom could be less than the number of templates - the matrix $C$ could have some small eigenvalues, leading to an overestimate of the number of degrees of freedom of the $\chi^2$ discriminator. In the ambiguity $\chi^2$, we remove these degrees of freedom by putting a lower cut-off on the eigenvalues, so we have a true estimate of the effectiveness of the $\chi^2$. In the autocorrelation $\chi^2$ \cite{Hanna,GSTLAL} the SNR time series is considered such that only the vectors along the time dimension are considered for consistency whereas in the ambiguity $\chi^2$, consistency is evaluated also in the other dimensions of the parameter space besides time, such as the masses.    
\par

We also indicated how the ambiguity $\chi^2$ discriminator could be generalized to the coherent multi-detector detection. For the case of uncorrelated noise between different detectors, the generalization is straight forward. We have not pursued this point here any further but may do so in the future.  
\par

Finally, we would like to mention that the vector bundle structure of the $\chi^2$ is more generally applicable.  In this paper we have discussed a specific signal model of the CBC. But if the signal model is different, either due to a modification of the CBC model used here or because of a completely different astrophysical source, the signal manifold $\P$ will differ, but the overall mathematical structure of the $\chi^2$, namely that of the vector bundle, will still remain the same. Therefore, the formalism for the $\chi^2$ presented here is valid in more general settings. 

\section*{Acknowledgments}

We would like to thank Reetika Dudi and Sanjit Mitra for helpful discussions, and Stanislav Babak for carefully reading the manuscript and making useful comments. One of us (SVD) would like to thank Bruce Allen and Badri Krishnan for discussions and visits to Albert Einstein Institute, Max Planck Institute for Gravitational Physics, Hannover, Germany. This work is supported in part by the Navajbai Ratan Tata Trust and NSF grant PHY-1506497. BG would like to thank University Grants Commission (UGC), India, for financial support as senior research fellow. All the numerical simulations are performed on IUCAA-LDG cluster (Sarathi). This paper has the LIGO document number P1700206.


\end{document}